\documentclass{article}
\usepackage[utf8]{inputenc}
\usepackage{refstyle}
\usepackage{amsmath}
\usepackage{amssymb}

\makeatletter

\AtBeginDocument{\providecommand\thmref[1]{\ref{thm:#1}}}
\AtBeginDocument{\providecommand\secref[1]{\ref{sec:#1}}}
\AtBeginDocument{\providecommand\subsecref[1]{\ref{subsec:#1}}}
\AtBeginDocument{\providecommand\Subsecref[1]{\ref{Subsec:#1}}}
\AtBeginDocument{\providecommand\conref[1]{\ref{con:#1}}}
\AtBeginDocument{\providecommand\defref[1]{\ref{def:#1}}}
\AtBeginDocument{\providecommand\Secref[1]{\ref{Sec:#1}}}
\AtBeginDocument{\providecommand\lemref[1]{\ref{lem:#1}}}
\AtBeginDocument{\providecommand\facref[1]{\ref{fac:#1}}}
\AtBeginDocument{\providecommand\proref[1]{\ref{pro:#1}}}
\AtBeginDocument{\providecommand\remref[1]{\ref{rem:#1}}}
\AtBeginDocument{\providecommand\corref[1]{\ref{cor:#1}}}
\AtBeginDocument{\providecommand\claref[1]{\ref{cla:#1}}}
\AtBeginDocument{\providecommand\enuref[1]{\ref{enu:#1}}}
\AtBeginDocument{\providecommand\eqref[1]{\ref{eq:#1}}}
\AtBeginDocument{\providecommand\Lemref[1]{\ref{Lem:#1}}}
\RS@ifundefined{subsecref}
  {\newref{subsec}{name = \RSsectxt}}
  {}
\RS@ifundefined{thmref}
  {\def\RSthmtxt{theorem~}\newref{thm}{name = \RSthmtxt}}
  {}
\RS@ifundefined{lemref}
  {\def\RSlemtxt{lemma~}\newref{lem}{name = \RSlemtxt}}
  {}

\usepackage{amsmath}
\usepackage{amsthm}
\usepackage{amssymb}
\usepackage{refstyle}

\theoremstyle{plain}
\newtheorem{theorem}{Theorem}[section]

\theoremstyle{plain}
\newtheorem{theorem-seq}{Theorem}

\newref{thm}{name=Theorem~, Name=Theorem~, names=Theorems~, Names=Theorems~}
\newref{subsec}{name=Section~, Name=Section~, names=Sections~, Names=Sections~}
\newref{sec}{name=Section~, Name=Section~, names=Sections~, Names=Sections~}

\newenvironment{myproof}[1][Proof.]{\par
	\pushQED{\qed}%
	\normalfont \topsep6\p@\@plus6\p@\relax
	\trivlist
		\item[\hskip\labelsep
		\bfseries
	#1]\ignorespaces
	}{%
	\popQED\endtrivlist\@endpefalse
}

\theoremstyle{plain}
\newtheorem*{rep@theorem}{\rep@title}

\newcommand{\newreptheorem}[2]{%
	\newenvironment{rep#1}[1]{%
	 \def\rep@title{#2 ##1}%
	 \begin{rep@theorem}}%
	 {\end{rep@theorem}}
}


\newcommand{\eqdef}{\stackrel{\rm{def}}{=}}

\newcommand{\N}{\mathbb{N}}

\newcommand{\R}{\mathbb{R}}

\ifx \C \undefined
\newcommand{\C}{\mathbb{C}}
\else
\renewcommand{\C}{\mathbb{C}}
\fi

\newcommand{\B}{\left\{ 0,1 \right \}}

\theoremstyle{definition}
\newtheorem{definition}[theorem]{Definition}
\newref{def}{name=Definition~, Name=Definition~, names=Definitions~, Names=Definition~}
\theoremstyle{definition}
\newtheorem{remark}[theorem]{Remark}
	\newref{rem}{name=Remark~, Name=Remark~, names=Remarks~, Names=Remarks~}
\theoremstyle{plain}    
\newtheorem{conjecture}[theorem]{Conjecture} 
	\newref{con}{name=Conjecture~, Name=Conjecture~, names=Conjectures~, Names=Conjectures~}
\theoremstyle{plain}    
\newtheorem{fact}[theorem]{Fact}
	\newref{fac}{name=Fact~, Name=Fact~, names=Facts~, Names=Facts~}
\theoremstyle{plain}    
\newtheorem{lemma}[theorem]{Lemma}  
	\newref{lem}{name=Lemma~, Name=Lemma~, names=Lemmas~, Names=Lemmas~}
\theoremstyle{definition}
\newtheorem*{rep@definition}{Definition \rep@ref}
\newenvironment{repdefinition}[1]{
 \def\rep@ref{#1}
 \begin{rep@definition}}
 {\end{rep@definition}}
\theoremstyle{plain}    
\newtheorem{corollary}[theorem]{Corollary}  
	\newref{cor}{name=Corollary~, Name=Corollary~, names=Corollarys~, Names=Corollaries~}
\newreptheorem{theorem}{Theorem}
\theoremstyle{plain}    
\newtheorem{proposition}[theorem]{Proposition} 
	\newref{pro}{name=Proposition~, Name=Proposition~, names=Propositions~, Names=Propositions~}
\newreptheorem{lemma}{Lemma}
\theoremstyle{plain}    
\newtheorem{claim}[theorem]{Claim}
	\newref{cla}{name=Claim~, Name=Claim~, names=Claims~, Names=Claims~}
\theoremstyle{definition}    
\newtheorem*{acknowledgement*}{Acknowledgement} 

\usepackage{hyperref}
\usepackage{amsthm}
\usepackage{fullpage}
\usepackage{color}

\makeatother

\begin{document}
\title{Query-to-Communication Lifting Using Low-Discrepancy Gadgets\thanks{This work subsumes an earlier work that appeared in ICALP 2019 \cite{CFKMP19}.
The earlier work proved our main result (\thmref{main}) only for
the special case where the gadget~$g$ is the inner product function,
while this work proves the result for the general case of all low-discrepancy
gadgets.}}
\author{Arkadev Chattopadhyay\thanks{School of Technology and Computer Science, Tata Institute of Fundamental
Research, Mumbai, India. \texttt{arkadev@tifr.res.in}.}\and Yuval Filmus\thanks{Technion Israel Institute of Technology, Haifa, Israel. \texttt{yuvalfi@cs.technion.ac.il}.
Taub Fellow --- supported by the Taub Foundations. The research was
funded by ISF grant 1337/16.}\and Sajin Koroth\thanks{School of Computing Science, Simon Fraser University, 8888 University
Drive, Burnaby, B.C., Canada V5A 1S6. This research was done while
Sajin Koroth was partially supported by the Israel Science Foundation
(grant No. 1445/16) and by the institutional postdoctoral program
of the University of Haifa.}\and Or Meir\thanks{Department of Computer Science, University of Haifa, Haifa 3498838,
Israel. \texttt{ormeir@cs.haifa.ac.il}. Partially supported by the
Israel Science Foundation (grant No. 1445/16).}\and Toniann Pitassi\thanks{Department of Computer Science, University of Toronto, Canada. \texttt{toni@cs.toronto.edu}.
Research supported by NSERC and. by NSF CCF grant 1900460}}
\maketitle
\begin{abstract}
Lifting theorems are theorems that relate the query complexity of
a function $f:\B^{n}\to\B$ to the communication complexity of the
composed function~$f\circ g^{n}$, for some ``gadget'' $g:\B^{b}\times\B^{b}\to\B$.
Such theorems allow transferring lower bounds from query complexity
to the communication complexity, and have seen numerous applications
in the recent years. In addition, such theorems can be viewed as a
strong generalization of a direct-sum theorem for the gadget~$g$.

We prove a new lifting theorem that works for all gadgets~$g$ that
have logarithmic length and exponentially-small discrepancy, for both
deterministic and randomized communication complexity. Thus, we significantly
increase the range of gadgets for which such lifting theorems hold.

Our result has two main motivations: First, allowing a larger variety
of gadgets may support more applications. In particular, our work
is the first to prove a randomized lifting theorem for logarithmic-size
gadgets, thus improving some applications of the theorem. Second,
our result can be seen as a strong generalization of a direct-sum
theorem for functions with low discrepancy.
\end{abstract}

\section{\label{sec:introduction}Introduction}

\subsection{\label{subsec:background}Background}

\global\long\def\dcc{D^{\mathrm{cc}}}%
\global\long\def\rcc{R^{\mathrm{cc}}}%
\global\long\def\ddt{D^{\mathrm{dt}}}%
\global\long\def\rdt{R^{\mathrm{dt}}}%
\global\long\def\Err{\mathrm{Err}}%
\global\long\def\disc{{\rm disc}}%
\global\long\def\cE{\mathcal{E}}%
\global\long\def\cV{\mathcal{V}}%
\global\long\def\D{\Lambda}%
\global\long\def\cX{\mathcal{X}}%
\global\long\def\cY{\mathcal{Y}}%
\global\long\def\fix{\mathrm{fix}}%
\global\long\def\free{\mathrm{free}}%
\global\long\def\D{\Lambda}%
\global\long\def\cI{\mathcal{I}}%
\global\long\def\cO{\mathcal{O}}%
\global\long\def\cS{\mathcal{S}}%
\global\long\def\b{\mathrm{bias}}%
\global\long\def\Hm{H_{\infty}}%
In this work, we prove new lifting theorems for a large family of
gadgets. Let $f\colon\{0,1\}^{n}\to\{0,1\}$ and $g\colon\{0,1\}^{b}\times\{0,1\}^{b}\to\{0,1\}$
be functions (where $g$ is referred to as a \emph{gadget}). The block-composed
function $f\circ g^{n}$ is the function that takes $n$ inputs $(x_{1},y_{1}),\ldots,(x_{n},y_{n})$
for $g$ and computes $f\circ g^{n}$ as, 
\[
f\circ g^{n}\left((x_{1},y_{1}),\ldots,(x_{n},y_{n}))=f(g(x_{1},y_{1}),g(x_{2},y_{2}),\ldots,g(x_{n},y_{n})\right).
\]
Lifting theorems are theorems that relate the communication complexity
of $f\circ g^{n}$ to the query complexity of $f$ and the communication
complexity of $g$.

More specifically, consider the following communication problem: Alice
gets $x_{1},\ldots,x_{n}$, Bob gets $y_{1},\ldots,y_{n}$, and they
wish to compute the output of $f\circ g^{n}$ on their inputs. The
natural protocol for doing so is the following: Alice and Bob jointly
\emph{simulate} a decision tree of optimal height for solving $f$.
Any time the tree queries the $i$-th bit, they compute $g$ on $(x_{i},y_{i})$
by invoking the best possible communication protocol for $g$. A \emph{lifting
theorem} is a theorem that says that this natural protocol is optimal.

We note that it is often desirable to consider the case where $f$
is a search problem with an arbitrary range rather than a boolean
function (see Section~\secref{preliminaries} for the definition
of search problems). Most of the known results, as well as the results
of this work, apply to this general case. However, for the simplicity
of presentation, we focus for now on the case where $f$ is a boolean
function.

\paragraph*{Applications of lifting theorems.}

One important reason for why lifting theorems are interesting is that
they create a connection between query complexity and communication
complexity. This connection, besides being interesting in its own
right, allows us to transfer lower bounds and separations from query
complexity (which is a relatively simple model) to communication complexity
(which is a significantly richer model).

In particular, the first result of this form, due to Raz and McKenzie~\cite{RM99},
proved a lifting theorem from \emph{deterministic} query complexity
to \emph{deterministic} communication complexity when $g$ is the
index function. They then used it to prove new lower bounds on communication
complexity by lifting query-complexity lower-bounds. More recently,
Göös, Pitassi and Watson~\cite{GPW15} applied that theorem to separate
the logarithm of the partition number and the deterministic communication
complexity of a function, resolving a long-standing open problem.
This too was done by proving such a separation in the setting of query
complexity and then lifting it to the setting of communication complexity.
This result stimulated a flurry of work on lifting theorems of various
kinds, such as: lifting for zero-communication protocols~\cite{GLMWZ},
round-preserving lifting theorems with applications to time-space
trade-offs for proof complexity \cite{RNV16}, deterministic lifting
theorems with other gadgets \cite{CKLM17,WYY17}, lifting theorems
from randomized query complexity to randomized communication complexity~\cite{GPW17},
lifting theorems for DAG-like protocols \cite{GGKS18} with applications
to monotone circuit lower bounds, lifting theorems for asymmetric
communication problems \cite{CKLM18} with applications to data-structures,
a lifting theorem for the EQUALITY gadget \cite{LM18}, lifting theorems
for XOR functions with applications to the log-rank conjecture~\cite{HHL18},
and lifting theorems for applications to monotone formula complexity,
monotone span programs, and proof complexity~\cite{DBLP:journals/siamcomp/GoosP18,RPRC16,PR17,PR18}.
There are also lifting theorems which lift more analytic properties
of the function like approximate degree due to Sherstov \cite{Sherstov11}
and independently due to Shi and Zhu \cite{SZ09}.

In almost all known lifting theorems, the function $f$ can be arbitrary
while $g$ is usually a specific function (e.g., the index function).
This raises the following natural question: for which choices of $g$
can we prove lifting theorems? This question is interesting because
usually the applications of lifting theorems work by reducing the
composed function $f\circ g^{n}$ to some other problem of interest,
and the choice of the gadget~$g$ affects the efficiency of such
reductions.

In particular, applications of lifting theorems often depend on the
size of the gadget, which is the length~$b$ of the input to $g$.
Both the deterministic lifting theorem of Raz and McKenzie~\cite{RM99}
and the randomized lifting theorem of Göös et al.~\cite{GPW17} use
a gadget of very large size (polynomial in $n$). Reducing the gadget
size to a constant would have many interesting applications.

In the deterministic setting, the gadget size was recently improved
to logarithmic by the independent works of~\cite{CKLM17} and~\cite{WYY17}.
Moreover, \cite{CKLM17,Koza18} showed the lifting to work for a class
of gadgets with a certain pseudorandom property rather than just a
single specific gadget. A gadget of logarithmic size was also obtained
earlier in lifting theorems for more specialized models, such as the
work of~\cite{GLMWZ}. However, the randomized lifting theorem of
Göös et al.~\cite{GPW17} seemed to work only with a specific gadget
of polynomial size.

In this work, we prove a lifting theorem for a large family of gadgets,
namely, all functions $g$ with logarithmic length and exponentially-small
discrepancy (see \subsecref{our-results} for details). Our theorem
holds both in the deterministic and the randomized setting. This allows
for a considerably larger variety of gadgets: in particular, our theorem
is the first lifting theorem in the randomized setting that uses logarithmic-size
gadgets, it allows lifting with the inner-product gadget (previously
known only in the deterministic setting \cite{CKLM17,WYY17}), and
it is also the first lifting theorem that shows that a random function
can be used as a gadget.

We would like to point out that, although we reduce the gadget size
to logarithmic in this work, it is not enough to obtain the interesting
applications a constant sized gadget would have yielded. Nevertheless,
our randomized lifting theorem still has some applications. For example,
our theorem can be used to simplify the lower bounds of Göös and Jayram
\cite{GJ16} on AND-OR trees and MAJORITY trees, since we can now
obtain them directly from the randomized query complexity lower bounds
rather than going through conical juntas. In addition, our theorem
can be used to derive the separation of randomized separation from
partition number (due to \cite{GJP015}) for functions with larger
complexity (compared to their input length).

\paragraph*{Lifting theorems as a generalization of direct-sum theorems.}

Lifting theorems can also be motivated from another angle, which is
particularly appealing in our case: lifting theorems can be viewed
as a generalization of direct-sum theorems. The direct-sum question
is a classical question in complexity theory, which asks whether performing
a task on $n$~independent inputs is $n$~times harder than performing
it on a single input. When specialized to the setting of communication
complexity, a direct-sum theorem is a theorem that says that the communication
complexity of a computing $g$ on $n$~independent inputs is about
$n$~times larger than the communication complexity of~$g$. A related
type of result, which is sometimes referred to as an ``XOR lemma'',
says that computing the XOR of the outputs of~$g$ on $n$~independent
inputs is about $m$~times larger than the communication complexity
of~$g$.

The direct-sum question for communication complexity has been raised
in~\cite{KRW91}, and has since attracted much attention. While we
do not have a general direct-sum theorem for all functions, many works
have proved direct-sum theorems and XOR lemmas for large families
of functions \cite{FKNN95,Shaltiel03,BPSW06,LSS08,Klauck10,BBCR10,BRWY13,B17}
as well as provided counterexamples \cite{FKNN95,GKR14,GKR16-boolean,GKR16-external}.

Now, observe that lifting theorems are natural generalizations of
direct-sum theorems and XOR lemmas: in particular, if we set $f$
to be the identity function or the parity function, we get a direct
sum theorem or an XOR lemma for $g$, respectively. More generally,
a lifting theorem says that the communication complexity of computing
any function~$f$ of the outputs of~$g$ on independent inputs is
larger than the complexity of~$g$ by a factor that depends on the
query complexity of~$f$. This is perhaps the strongest and most
natural ``direct-sum-like theorem'' for~$g$ that one could hope
for.

From this perspective, it is natural to ask which functions~$g$
admit such a strong theorem. The previous works of \cite{RM99,GPW17}
can be viewed as establishing this theorem only for the index function.
The work of \cite{CKLM17,Koza18} have made further progress, establishing
this theorem for a class of functions with satisfy certain hitting
property. However, the latter property is somewhat non-standard and
ad hoc, and their theorem holds only in the deterministic setting.
In this work, we establish such theorems for all functions~$g$ with
low discrepancy, which is a standard and well-studied complexity measure,
and we do so in both the deterministic and the randomized setting.

\subsection{\label{subsec:our-results}Our results}

In this work, we prove lifting theorems for gadgets of low-discrepancy.
In what follows, we denote by $\ddt$ and $\dcc$ the deterministic
query complexity and communication complexity of a task respectively,
and by $\rdt_{\varepsilon}$ and $\rcc_{\varepsilon}$ the randomized
query complexity and (public-coin) communication complexity with error
probability~$\varepsilon$ respectively. Given a search problem $\cS$
and a gadget $g:\B^{b}\times\B^{b}\to\B$, it is easy to see that
\begin{align*}
\dcc(\cS\circ g^{n}) & =O\left(\ddt(\cS)\cdot b\right)\\
\rcc_{\varepsilon}(\cS\circ g^{n}) & =O\left(\rdt(\cS)\cdot b\right).
\end{align*}
This upper bound is proved using the simple protocol discussed earlier:
the party simulates the optimal decision tree for~$\cS$, and whenever
a query is made, the parties compute~$g$ on the corresponding input
in order to answer the query (which can be done by communicating at
most $b+1$ bits). Our main result says that when $g$~has low discrepancy
and $b$ is at least logarithmic, that this upper bound is roughly
tight. In order to state this result, we first recall the definition
of discrepancy.
\begin{definition}
\label{def:discrepancy}Let $\D$ be a finite set, let $g:\D\times\D\to\B$
be a function, and let $U,V$ be independent random variables that
are uniformly distributed over~$\D$. Given a combinatorial rectangle
$R\subseteq\D\times\D$, the \emph{discrepancy of~$g$ with respect
to~$R$}, denoted $\disc_{R}(g)$, is defined as follows:
\[
\disc_{R}(g)=\left|\Pr\left[g(U,V)=0\text{ and }(U,V)\in R\right]-\Pr\left[g(U,V)=1\text{ and }(U,V)\in R\right]\right|.
\]
The \emph{discrepancy of~$g$}, denoted $\disc(g)$, is defined as
the maximum of $\disc_{R}(g)$ over all combinatorial rectangles~$R\subseteq\D\times\D$.
\end{definition}
Discrepancy is a useful measure for the complexity of~$g$, and in
particular, it is well-known that for $\varepsilon>0$:
\[
\dcc(g)\ge\rcc_{\varepsilon}(g)\ge\log\frac{1-2\cdot\varepsilon}{\disc(g)}
\]
(see, e.g., \cite{KN97}). We now state our main result.
\begin{theorem}[Main theorem]
\label{thm:main}For every $\eta>0$ there exists $c=O(\frac{1}{\eta^{2}}\cdot\log\frac{1}{\eta})$
such that the following holds: Let $\cS$ be a search problem that
takes inputs from~$\B^{n}$, and let $g:\B^{b}\times\B^{b}\to\B$
be an arbitrary function such that $\disc(g)\le2^{-\eta\cdot b}$
and such that $b\ge c\cdot\log n$. Then 
\[
\dcc(\cS\circ g^{n})=\Omega\left(\ddt(\cS)\cdot b\right),
\]
and for every $\varepsilon>0$ it holds that
\[
\rcc_{\varepsilon}(\cS\circ g^{n})=\Omega\left(\left(\rdt_{\varepsilon'}(\cS)-O(1)\right)\cdot b\right),
\]
where $\varepsilon'=\varepsilon+2^{-\eta\cdot b/8}$.
\end{theorem}
We note that our results are in fact more general, and preserve the
round complexity of~$\cS$ among other things. See Sections \ref{sec:deterministic-lifting}
and \ref{sec:randomized-lifting} for more details.
\begin{remark}
Note that our main theorem can be applied to a random function~$g:\B^{b}\times\B^{b}\to\B$,
since such a function has a very low discrepancy. As noted above,
we believe that our theorem is the first theorem to allow the gadget
to be a random function.
\end{remark}

\paragraph*{Unifying deterministic and randomized lifting theorems.}

The existing proofs of deterministic lifting theorems and randomized
lifting theorems are quite different. While both proofs rely on information-theoretic
arguments, they measure information in different ways. In particular,
while the randomized lifting theorem of \cite{GPW17} (following~\cite{GLMWZ})
measures information using min-entropy, the deterministic lifting
theorems of~\cite{RM99,GPW15,CKLM17,WYY17} (following \cite{EIRS01})
measure information using a notion known as \emph{thickness} (with
\cite{GGKS18} being a notable exception). A natural direction of
further research is to investigate if these disparate techniques can
be unified. Indeed, a related question was raised by \cite{GLMWZ},
who asked if min-entropy based techniques could be used to prove (or
simplify the existing proof of) Raz--McKenzie style deterministic
lifting theorems.

Our work answers this question affirmatively: we prove both the deterministic
and randomized lifting theorems using the same strategy. In particular,
both proofs measure information using min-entropy. In doing so, we
unify both lifting theorems under the same framework.

\subsection{\label{subsec:our-techniques}Our techniques}

We turn to describe the high-level ideas that underlie the proof of
our main theorem. Following the previous works, we use a ``simulation
argument'': We show that given a protocol~$\Pi$ that solves $\cS\circ g^{n}$
with communication complexity~$C$, we can construct a decision tree~$T$
that solves~$\cS$ with query complexity~$O(\frac{C}{b})$. The
decision tree~$T$ works by simulating the action of the protocol~$\Pi$
(hence the name ``simulation argument''). We now describe this simulation
in more detail, following the presentation of~\cite{GPW17}.

\paragraph*{The simulation argument.}

For simplicity of notation, let us denote $\D=\B^{b}$, so $g$ is
a function from a ``block'' in $\D\times\D$ to $\B$. Let $G=g^{n}:\D^{n}\times\D^{n}\to\B^{n}$
be the function that takes $n$~disjoint blocks and computes the
outputs of~$g$ on all of them. We assume that we have a protocol~$\Pi$
that solves $\cS\circ G$ with complexity~$C$, and would like to
construct a decision tree~$T$ that solves~$\cS$ with complexity~$O(\frac{C}{b})$.
The basic idea is that given an input~$z\in\B^{n}$, the tree~$T$
simulates the action of~$\Pi$ on the random inputs $(X,Y)$ that
are uniformly distributed over $G^{-1}(z)$. Clearly, it holds that
$\cS\circ G(X,Y)=\cS(z)$, so this simulation, if done right, outputs
the correct answer.

The core issue in implementing such a simulation is the following
question: how can $T$ simulate the action of $\Pi$ on $(X,Y)\in G^{-1}(z)$
without knowing~$z$? The answer is that as long as the protocol~$\Pi$
has transmitted less than $\varepsilon\cdot b$~bits of information
about every block $(X_{i},Y_{i})$ (for some specific~$\varepsilon>0$),
the distribution of $(X,Y)$ is similar to the uniform distribution
in a certain sense (that will be formalized soon). Thus, the tree~$T$
can pretend that $(X,Y)$ are distributed uniformly, and simulate
the action of~$\Pi$ on such inputs, which can be done without knowing~$z$.

This idea can be implemented as long as the protocol has transmitted
less than $\varepsilon\cdot b$~bits of information about every block
$(X_{i},Y_{i})$. However, at some point, the protocol may transmit
more than $\varepsilon\cdot b$ bits of information about some blocks.
Let $I\subseteq\left[n\right]$ denote the set of these blocks. At
this point, it is no longer true that the distribution of $(X,Y)$
is similar to the uniform distribution. However, it can be shown that
the distribution of~$(X,Y)$ is similar to the uniform distribution
\emph{conditioned on $g^{I}(X_{I},Y_{I})=z_{I}$}. Thus, the tree~$T$
queries the bits in~$z_{I}$, and can now continue the simulation
of $\Pi$ on $(X,Y)\in G^{-1}(z)$ by pretending that $(X,Y)$ are
distributed uniformly conditioned on~$g^{I}(X_{I},Y_{I})=z_{I}$.
The tree proceeds in this way, adding blocks to~$I$ as necessary,
until the protocol~$\Pi$ ends, at which point~$T$ halts and outputs
the same output as~$\Pi$.

It remains to show that the query complexity of~$T$ is at most~$O(\frac{C}{b})$.
To this end, observe that the query complexity of~$T$ is exactly
the size of the set~$I$ at the end of the simulation. Moreover,
recall that the set~$I$ is the set of blocks on which the protocol
transmitted at least $\varepsilon\cdot b$ bits of information. Hence,
at any given point, the protocol must have transmitted at least $\varepsilon\cdot b\cdot\left|I\right|$
bits. On the other hand, we know by assumption that the protocol never
transmitted more than~$C$ bits. This implies that $\varepsilon\cdot b\cdot\left|I\right|\le C$
and therefore the query complexity of the tree~$T$ is at most $\left|I\right|\le\frac{C}{\varepsilon\cdot b}=O(\frac{C}{b})$.
This concludes the argument.

\paragraph*{Our contribution.}

In order to implement the foregoing simulation argument, there are
two technical issues that need to be addressed and are relevant at
this point:
\begin{itemize}
\item \textbf{The uniform marginals issue:} In the above description, we
argued that as long the protocol has not transmitted too much information,
the distribution of $(X,Y)$ is ``similar to the uniform distribution''.
The question is how do we formalize this idea. This issue was dealt
with implicitly in several works in the lifting literature since~\cite{RM99},
and was made explicit in \cite{GPW17} as the ``uniform marginals
lemma'': if every set of blocks in~$(X,Y)$ has sufficient min-entropy,
then each of the marginals $X,Y$ on its own is close to the uniform
distribution. In \cite{GPW17}, they proved this lemma for the case
where $g$~is the index function, and in \cite{GLMWZ} a very similar
lemma was proved for the case where $g$ is the inner product function.
\item \textbf{The conditioning issue:} As we described above, when the protocol
transmits too much information about a set of blocks $I\subseteq\left[n\right]$,
the tree~$T$ queries $z_{I}$ and conditions the distribution of~$(X,Y)$
on the event that $g^{I}(X_{I},Y_{I})=z_{I}$. In principle, this
conditioning may reveal information on $(X_{\left[n\right]-I},Y_{\left[n\right]-I})$,
which might reduce their min-entropy and ruin their uniform-marginals
property. In order for the simulation argument to work, one needs
to show that this cannot happen, and the conditioning will never reveal
too much information about $(X_{\left[n\right]-I},Y_{\left[n\right]-I})$.\\
In the works of \cite{RM99,WYY17,GPW17} this issue was handled by
arguments that are tailored to the index and inner product functions.
The work of \cite{CKLM17} gave this issue a more general treatment,
by identifying an abstract property of~$g$ that prevents the conditioning
from revealing too much information. However, as discussed above,
this abstract property is somewhat ad hoc, and only works for deterministic
simulation.
\end{itemize}
Our contribution is dealing with both issues in the general setting
where $g$ is an arbitrary low-discrepancy gadget. In order to deal
with the first issue, we prove a ``uniform marginals'' lemma for
such gadgets~$g$: this is relatively easy, since the proof of~\cite{GLMWZ}
for the inner product gadget turns out to generalize in a straightforward
way to arbitrary low-discrepancy gadgets.

The core of this work is in dealing with the conditioning issue. Our
main technical lemma that says that as long as every set of blocks
in~$(X,Y)$ has sufficient min-entropy, there are only few possible
values of $X,Y$ that are ``dangerous'' (in the sense that they
may lead the conditioning to leak too much information). We now modify
the simulation such that it discards these dangerous values before
performing the conditioning. Since there are only few of those dangerous
values, discarding them does not reveal too much information on~$X$
and~$Y$, and the simulation can proceed as before.

\subsection{Open problems}

The main question that arises from this work is how much more general
the gadget~$g$ can be? As was discussed in \Subsecref{background},
lifting theorems can be viewed as a generalization of direct-sum theorems.
In the setting of randomized communication complexity, it is known
that the ``ability of~$g$ to admit a direct-sum theorem'' is characterized
exactly by a complexity measure called the \emph{information cost}
of~$g$ (denoted~$\boldsymbol{IC}(g)$). In particular, the complexity
of computing a function~$g$ on $n$~independent copies is $\approx n\cdot\boldsymbol{IC}(g)$.
\cite{BBCR10,BR14,B17}. This leads to the natural conjecture that
a lifting theorem should hold for every gadget~$g$ that has sufficiently
high information cost.
\begin{conjecture}
\label{con:information-lifting}There exists a constant $c>0$ such
that the following holds. Let $\cS$ be any search problem that takes
inputs from~$\B^{n}$, and let $g:\B^{b}\times\B^{b}\to\B$ be an
arbitrary function such that $\boldsymbol{IC}(g)\ge c\cdot\log n$.
Then 
\[
\rcc_{\varepsilon}(\cS\circ g^{n})=\Omega\left(\rdt_{\varepsilon'}(\cS)\cdot\boldsymbol{IC}(g)\right).
\]
\end{conjecture}
\noindent Proving this conjecture would give us a nearly-complete
understanding of the lifting phenomenon which, in addition to being
interesting in its own right, would likely lead to many applications.
In particular, this conjecture implies our result, since it is known
that $\log\frac{1}{\disc(g)}$ (roughly) lower bounds the information
cost of~$g$ \cite{KLLRX15}.

\conref{information-lifting} is quite ambitious. As intermediate
goals, one could attempt to prove such a lifting theorem for other
complexity measures that are stronger than discrepancy and weaker
than information cost (see \cite{JK10,KLLRX15} for several measures
of this kind). To begin with, one could consider the well-known corruption
bound of~\cite{Y83,BFS86,R92}: could we prove a lifting theorem
for an arbitrary gadget~$g$ that has a low corruption bound? A particularly
interesting example for such a gadget is the disjointness function
--- indeed, proving a lifting theorem for the disjointness gadget
would be interesting in its own right and would likely have applications,
in addition to being a step toward \conref{information-lifting}.

An even more modest intermediate goal is to gain better understanding
of lifting theorems with respect to discrepancy. For starters, our
result only holds\footnote{More accurately, our result can be applied to gadgets with larger
discrepancy, but then the gadget size has to be larger than logarithmic.} for gadgets whose discrepancy is exponentially vanishing in the gadget
size. Can we prove a lifting theorem for gadgets~$g$ with larger
discrepancy? In particular, since the randomized communication complexity
of~$g$ is lower bounded by $\Omega(\log\frac{1}{\disc(g)})$, the
following conjecture comes to mind.
\begin{conjecture}
\label{con:lifting-discrepancy}There exists a constant $c>0$ such
that the following holds. Let $\cS$ be any search problem that takes
inputs from~$\B^{n}$, and let $g:\B^{b}\times\B^{b}\to\B$ be an
arbitrary function such that $\log\frac{1}{\disc(g)}\ge c\cdot\log n$.
Then 
\[
\rcc_{\varepsilon}(\cS\circ g^{n})=\Omega\left(\rdt_{\varepsilon'}(\cS)\cdot\log\frac{1}{\disc(g)}\right).
\]
\end{conjecture}
\noindent Another interesting direction is to consider discrepancy
with respect to other distributions. The definition of discrepancy
we gave above (\defref{discrepancy}) is a special case of a more
general definition, in which the random variables $(U,V)$ are distributed
according to some fixed distribution~$\mu$ over~$\D\times\D$.
Thus, our result works only when $\mu$ is the uniform distribution.
Can we prove a lifting theorem that holds for an arbitrary choice
of~$\mu$? While we have not verified it, we believe that our proof
can yield a lifting theorem that works whenever $\mu$~is a product
distribution (after some natural adaptations). However, proving such
a lifting theorem for non-product distributions seems to require new
ideas. We note that direct-sum theorems for discrepancy have been
proved by~\cite{Shaltiel03,LSS08}, and proving \conref{information-lifting}
(and extending it to an arbitrary distribution~$\mu$) seems like
a natural extension of their results.

Yet another interesting direction is to consider the lifting analogue
of strong direct product theorems. Such theorems say that when we
compute $g$ on $n$~independent inputs, then not only that the communication
complexity increases by a factor of~$n$, but the success probability
also drops exponentially in~$n$ (see, e.g., \cite{Shaltiel03,Klauck10,D12,BRWY13}).
A plausible analogue for lifting theorems is to conjecture that the
success probability of computing $\cS\circ g^{n}$ drops exponentially
in the query complexity of~$\cS$. It would be interesting to see
a result along these lines.

Finally, there remains major open problem of the lifting literature
to prove a lifting theorem that uses gadgets of constant size.\smallskip{}

\textbf{Organization of the paper.} In \Secref{preliminaries}, we
provide the required preliminaries. In \secref{lifting-machinery},
we set up the lifting machinery that is used in both the deterministic
and the randomized lifting results, including our ``uniform marginals
lemma'' and our main technical lemma. We prove the deterministic
part of our main theorem in \secref{deterministic-lifting}, and the
randomized part of our main theorem in \secref{randomized-lifting}.

\section{\label{sec:preliminaries}Preliminaries}

We assume familiarity with the basic definitions of communication
complexity (see, e.g., \cite{KN97}). For any $n\in\N$, we denote
$\left[n\right]\eqdef\left\{ 1,\ldots,n\right\} $. Given a boolean
random variable~$V$, we denote the bias of~$V$ by
\[
\b(V)\eqdef\left|\Pr\left[V=0\right]-\Pr\left[V=1\right]\right|.
\]
Given an alphabet~$\D$ and a set~$I\subseteq\left[n\right]$, we
denote by $\D^{I}$ the set of strings of length~$\left|I\right|$
which are indexed by~$I$. Given a string~$x\in\D^{n}$ and a set
$I\subseteq\left[n\right]$, we denote by $x_{I}$ the projection
of~$x$ to the coordinates in~$I$ (in particular, $x_{\emptyset}$
is defined to be the empty string). Given a boolean function~$g:\cX\times\cY\to\B$
and a set $I\subseteq\left[n\right]$, we denote by $g^{I}:\cX^{I}\times\cY^{I}\to\B^{I}$
the function that takes as inputs $\left|I\right|$ pairs from~$\cX\times\cY$
that are indexed by~$I$, and outputs the string in~$\B^{I}$ whose
$i$-th bit is the output of $g$ on the $i$-th pair. In particular,
we denote $g^{n}\eqdef g^{\left[n\right]}$, so the $g^{n}$ takes
as inputs $x\in\cX^{n},y\in\cY^{n}$ and outputs the binary string
\[
g^{n}(x,y)\eqdef\left(g(x_{1},y_{1}),\ldots,g(x_{n},y_{n})\right).
\]
For every $I\subseteq\left[n\right]$, we denote by $g^{\oplus I}:\cX^{I}\times\cY^{I}\to\B$
the function that given $x\in\cX^{I}$ and~$y\in\cY^{I}$, outputs
the parity of the string $g^{I}(x,y)$.

\paragraph*{Search problems.}

Given a finite set of inputs $\cI$ and a finite set of outputs~$\cO$,
a \emph{search problem}~$\cS$ is a relation between $\cI$ and~$\cO$.
Given $z\in\cI$, we denote by $\cS(z)$ the set of outputs $o\in\cO$
such that $(z,o)\in\cS$. Without loss of generality, we may assume
that $\cS(z)$ is always non-empty, since otherwise we can set $\cS(z)=\left\{ \bot\right\} $
where $\bot$ is some special failure symbol that does not belong
to~$\cO$.

Intuitively, a search problem~$\cS$ represents the following task:
given an input $z\in\cI$, find a solution $o\in\cS(z)$. In particular,
if $\cI=\cX\times\cY$ for some finite sets $\cX,\cY$, we say that
a deterministic protocol~$\Pi$ \emph{solves}~$\cS$ if for every
input $(x,y)\in\cI$, the output of $\Pi$ is in $\cS(x,y)$. We say
that a randomized protocol~$\Pi$ \emph{solves}~$\cS$ \emph{with
error}~$\varepsilon$ if for every input $(x,y)\in\cI$, the output
of~$\Pi$ is in $\cS(x,y)$ with probability at least~$1-\varepsilon$.

We denote the deterministic communication complexity of a search problem~$\cS$
with $\dcc(\cS)$. Given $\varepsilon>0$, we denote by $\rcc_{\varepsilon}(\cS)$
the randomized (public-coin) communication complexity of~$\cS$ with
error~$\varepsilon$ (i.e., the minimum worst-case complexity of
a randomized protocol that solves~$\cS$ with error~$\varepsilon$).

Given a search problem~$\cS\subseteq\B^{n}\times\cO$, we denote
by $\cS\circ g^{n}\subseteq(\cX^{n}\times\cY^{n})\times\cO$ the search
problem that satisfies for every $x\in\cX^{n}$ and $y\in\cY^{n}$
that $\cS\circ g^{n}(x,y)=\cS(g^{n}(x,y))$.

\subsection{Decision trees}

Informally, a decision tree is an algorithm that solves a search problem~$\cS\subseteq\B^{n}\times\cO$
by querying the individual bits of its input. The tree is computationally
unbounded, and its complexity is measured by the number of bits it
queried.

Formally, a \emph{deterministic decision tree}~$T$ from $\B^{n}$
to~$\cO$ is a binary tree in which every internal node is labeled
with a coordinate in~$\left[n\right]$ (which represents a query),
every edge is labeled by a bit (which represents the answer to the
query), and every leaf is labeled by an output in~$\cO$. Such a
tree computes a function from~$\B^{n}$ to~$\cO$ in the natural
way, and with a slight abuse of notation, we denote this function
also as~$T$. The \emph{query complexity} of~$T$ is the depth of
the tree. We say that a tree~$T$ solves a search problem $\cS\subseteq\B^{n}\times\cO$
if for every $z\in\B^{n}$ it holds that $T(z)\in\cS(z)$. The \emph{deterministic
query complexity of}~$\cS$, denoted $\ddt(\cS)$, is the minimal
query complexity of a decision tree that solves~$\cS$.

A \emph{randomized decision tree}~$T$ is a random variable that
takes deterministic decision trees as values. The \emph{query complexity}
of~$T$ is the maximal depth of a tree in the support of~$T$. We
say that $T$ \emph{solves a search problem $\cS\subseteq\B^{n}\times\cO$
with error~$\varepsilon$} if for every $z\in\B^{n}$ it holds that
\[
\Pr\left[T(z)\in\cS(z)\right]\ge1-\varepsilon.
\]
The \emph{randomized query complexity of~$\cS$ with error~$\varepsilon$},
denoted $\rdt_{\varepsilon}$, is the minimal query complexity of
a randomized decision tree that solves~$\cS$ with error~$\varepsilon$.
Again, when we omit $\varepsilon$, it is assumed to be~$\frac{1}{3}$.

\subsubsection{Parallel decision-trees}

Our lifting theorems have the property that they preserve the round
complexity of protocols, which is useful for some applications~\cite{RNV16}.
In order to define this property, we need a notion of a decision tree
that has an analogue of ``round complexity''. Such a notion, due
to~\cite{V75}, is called a \emph{parallel decision tree}. Informally,
a parallel decision tree is a decision tree that works in ``rounds'',
where in each round multiple queries are issued simultaneously. The
``round complexity'' of the tree is the number of rounds, whereas
the query complexity is the total number of queries issued.

Formally, a \emph{deterministic parallel decision tree~$T$} from
$\B^{n}$ to~$\cO$ is a rooted tree in which every internal node
is labeled with a set $I\subseteq\left[n\right]$ (representing the
queries issued simultaneously at this round) and has degree $2^{\left|I\right|}$.
The edges going out of such a node are labeled with all the possible
assignments in~$\B^{I}$, and the every leaf is labeled by some output~$o\in\cO$.
As before, such a tree naturally computes a function that is denoted
by~$T$, and it solves a search problem $\cS\subseteq\B^{n}\times\cO$
if $T(z)\in\cS(z)$ for all $z\in\B^{n}$. The depth of such a tree
is now the analogue of the number of rounds in a protocol. The \emph{query
complexity} of~$T$ is defined as the maximum, over all leaves~$\ell$,
of the sum of the sizes of the sets~$I$ that are labeling the vertices
on the path from the root to~$\ell$. A \emph{randomized parallel
decision tree} is defined analogously to the definition of randomized
decision trees above.

\subsection{Fourier analysis}

Given a set~$S\subseteq\left[m\right]$, the \emph{character}~$\chi_{S}$
is the function from~$\B^{m}$ to~$\R$ that is defined by
\[
\chi_{S}(z)\eqdef\left(-1\right)^{\bigoplus_{i\in S}z_{i}}.
\]
Here, if $S=\emptyset$ then we define $\bigoplus_{i\in S}z_{i}=0$.
Given a function~$f:\B^{m}\to\R$, its \emph{Fourier coefficient}~$\hat{f}(S)$
is defined as
\[
\hat{f}(S)\eqdef\frac{1}{2^{m}}\sum_{z\in\B^{m}}f(z)\cdot\chi_{S}(z).
\]
It is a standard fact of Fourier analysis that $f$ can be written
as
\begin{equation}
f(z)=\sum_{S\subseteq\left[m\right]}\hat{f}(S)\cdot\chi_{S}(z).\label{eq:fourier-inversion-formula}
\end{equation}
We have the following useful observation.
\begin{fact}
\label{fac:bias-vs-fourier}Let $Z$ be a random variable taking values
in~$\B^{m}$, and let $\mu:\B^{m}\to\R$ be its density function.
Then, for every set~$S\subseteq\left[m\right]$ it holds that
\[
\left|\hat{\mu}(S)\right|=2^{-m}\cdot\b(\bigoplus_{i\in S}Z_{i}).
\]
In particular, $\hat{\mu}(\emptyset)=2^{-m}$.
\end{fact}
\begin{myproof}
Let $S\subseteq\left[m\right]$. It holds that
\begin{align*}
\left|\hat{\mu}(S)\right| & =2^{-m}\cdot\left|\sum_{z\in\B^{m}}\mu(z)\cdot\chi_{S}(z)\right|\\
 & =2^{-m}\cdot\left|\sum_{z\in\B^{m}}\mu(z)\cdot\left(-1\right)^{\bigoplus_{i\in S}z_{i}}\right|\\
 & =2^{-m}\cdot\left|\sum_{z\in\B^{m}:\bigoplus_{i\in S}z_{i}=0}\mu(z)-\sum_{z\in\B^{m}:\bigoplus_{i\in S}z_{i}=1}\mu(z)\right|\\
 & =2^{-m}\cdot\left|\Pr\left[\bigoplus_{i\in S}Z_{i}=0\right]-\Pr\left[\bigoplus_{i\in S}Z_{i}=1\right]\right|\\
 & =2^{-m}\cdot\b(\bigoplus_{i\in S}Z_{i}),
\end{align*}
as required. The ``in particular'' part follows by noting that in
the case of~$S=\emptyset$, the character $\chi_{S}$ is the constant
function~$1$, and recalling that the sum of $\mu(z)$ over all $z$'s
is~$1$.
\end{myproof}

\subsection{Probability}

Given two distributions $\mu_{1},\mu_{2}$ over a finite sample space~$\Omega$,
the \emph{statistical distance} (or \emph{total variation distance})
between $\mu_{1}$ and~$\mu_{2}$ is
\[
\left|\mu_{1}-\mu_{2}\right|=\max_{\cE\subseteq\Omega}\left\{ \left|\mu_{1}(\cE)-\mu_{2}(\cE)\right|\right\} .
\]
It is not hard to see that the maximum is attained when $\cE$ consists
of all the values~$\omega\in\Omega$ such that $\mu_{1}(\omega)>\mu_{2}(\omega)$.
We say that $\mu$ and $\mu_{2}$ are $\varepsilon$\emph{-close}
if $\left|\mu-\mu_{2}\right|\le\varepsilon$. The \emph{min-entropy
}of a random variable~$X$, denoted~$\Hm(X)$, is the largest number~$k\in\R$
such that for every value~$x$ it holds that
\[
\Pr\left[X=x\right]\le2^{-k}.
\]
Min-entropy has the following easy-to-prove properties.
\begin{fact}
\label{fac:min-entropy-conditioning}Let $X$ be a random variable
and let $\cE$ be an event. Then, $\Hm(X\texttt{\ensuremath{\mid}}\cE)\ge\Hm(X)-\log\frac{1}{\Pr\left[\cE\right]}$.
\end{fact}
\begin{fact}
\label{fac:min-entrpoy-projecting}Let $X_{1},X_{2}$ be random variables
taking values from sets~$\cX_{1},\cX_{2}$ respectively. Then, $\Hm(X_{1})\ge H_{\infty}(X_{1},X_{2})-\log\left|\cX_{2}\right|$.
\end{fact}
\noindent We say that a distribution is $k$-flat if it is uniformly
distributed over a subset of the sample space of size at least~$2^{k}$.
The following standard fact is useful.
\begin{fact}
\label{fac:flat-distributions}If a random variable $X$ has min-entropy~$k$,
then its distribution is a convex combination of $k$-flat distributions.
\end{fact}

\subsubsection{Vazirani's Lemma}

Vazirani's lemma is a useful result which says that a random string
is close to being uniformly distributed if the XOR of every set of
bits in the string has a small bias. We use the following variant
of the lemma due to~\cite{GLMWZ}.
\begin{lemma}[\cite{GLMWZ}]
\label{lem:vazirani}Let $\varepsilon>0$, and let $Z$ be a random
variable taking values in~$\B^{m}$. If for every non-empty set $S\subseteq\left[m\right]$
it holds that 
\begin{equation}
\b(\bigoplus_{i\in S}Z_{i})\le\varepsilon\cdot\left(2\cdot m\right)^{-\left|S\right|}\label{eq:bias}
\end{equation}
then for every~$z\in\B^{m}$ it holds that 
\[
\left(1-\varepsilon\right)\cdot\frac{1}{2^{m}}\le\Pr\left[Z=z\right]\le\left(1+\varepsilon\right)\cdot\frac{1}{2^{m}}.
\]
\end{lemma}
\begin{myproof}
Let $\mu:\B^{m}\to\R$ be the density function of~$Z$, and let $z\in\B^{m}$.
By Equation~\ref{eq:fourier-inversion-formula} it holds that
\begin{align*}
\left|\mu(z)-2^{-m}\right| & =\left|\sum_{S\subseteq\left[m\right]}\hat{\mu}(S)\cdot\chi_{S}(z)-2^{-m}\right|\\
\text{(\ensuremath{\hat{\mu}(\emptyset)=2^{-m}} by Fact \ref{fac:bias-vs-fourier})} & =\left|\sum_{S\subseteq\left[m\right]:S\ne\emptyset}\hat{\mu}(S)\cdot\chi_{S}(z)\right|\\
\text{(Since \ensuremath{\left|\chi_{S}(z)\right|} is always\,\ensuremath{1})} & \le\sum_{S\subseteq\left[m\right]:S\ne\emptyset}\left|\hat{\mu}(S)\right|\\
\text{(Fact \ref{fac:bias-vs-fourier})} & =2^{-m}\cdot\sum_{S\subseteq\left[m\right]:S\ne\emptyset}\b(\bigoplus_{i\in S}Z_{i})\\
\text{(Inequality\,\ref{eq:bias})} & \le2^{-m}\cdot\sum_{S\subseteq\left[m\right]:S\ne\emptyset}\varepsilon\cdot\left(2\cdot m\right)^{-\left|S\right|}\\
 & =\varepsilon\cdot2^{-m}\cdot\sum_{i=1}^{m}\binom{m}{i}\cdot\left(2\cdot m\right)^{-i}\\
 & \le\varepsilon\cdot2^{-m}\cdot\sum_{i=1}^{m}m^{i}\cdot\left(2\cdot m\right)^{-i}\\
 & \le\varepsilon\cdot2^{-m}\cdot\sum_{i=1}^{m}2^{-i}\\
 & \le\varepsilon\cdot2^{-m},
\end{align*}
as required.
\end{myproof}
\lemref{vazirani} says that if the bias of~$\bigoplus_{i\in S}Z_{i}$
is small for every~$S$, then $Z$ is close to being uniformly distributed.
It turns out that if the latter assumption holds only for large sets~$S$,
we can still deduce something useful, namely, that the min-entropy
of~$Z$ is high.
\begin{lemma}
\label{lem:vazirani-min-entropy}Let $t\in\N$ be such that $t\ge1$,
and let $Z$ be a random variable taking values in~$\B^{m}$. If
for every set $S\subseteq\left[m\right]$ such that $\left|S\right|\ge t$
it holds that 
\[
\b(\bigoplus_{i\in S}Z_{i})\le\left(2\cdot m\right)^{-\left|S\right|},
\]
then, $\Hm(Z)\ge m-t\log m-1$.
\end{lemma}
\begin{myproof}
Observe that if $m=1$ then the bound holds vacuously, so we may assume
that $m\ge2$. Let $\mu:\B^{m}\to\R$ be the density function of~$Z$,
and let $z\in\B^{m}$. By Equality~\ref{eq:fourier-inversion-formula}
it holds that
\begin{align*}
\mu(z) & =\sum_{S\subseteq\left[m\right]}\hat{\mu}(S)\cdot\chi_{S}(z)\\
\text{(Since \ensuremath{\left|\chi_{S}(z)\right|} is always\,\ensuremath{1})} & \le\sum_{S\subseteq\left[m\right]}\left|\hat{\mu}(S)\right|\\
\text{(Fact \ref{fac:bias-vs-fourier})} & \le2^{-m}\cdot\sum_{S\subseteq\left[m\right]}\b(\oplus_{i\in S}Z_{i})\\
 & \le2^{-m}\cdot\left(\sum_{S\subseteq\left[m\right]:\left|S\right|<t}\b(\oplus_{i\in S}Z_{i})+\sum_{S\subseteq\left[m\right]:\left|S\right|\ge t}\b(\oplus_{i\in S}Z_{i})\right)
\end{align*}
We now bound each of the two terms separately. The term for sets~$S$
whose size is at least~$t$ can be upper bounded by $1$ using exactly
the same calculation as in the proof of \lemref{vazirani}. In order
to upper bound the term for sets whose size is less than~$t$, observe
that $\b(\oplus_{i\in S}Z_{i})\le1$ for every~$S\subseteq\left[m\right]$
and and therefore
\begin{align*}
\sum_{S\subseteq\left[m\right]:\left|S\right|<t}\b(\oplus_{i\in S}Z_{i}) & \le\sum_{i=0}^{t-1}\binom{m}{i}\\
 & \le\sum_{i=0}^{t-1}m^{i}\\
 & =\frac{m^{t}-1}{m-1}\\
\text{(Since \ensuremath{m\ge2})} & \le m^{t}-1.
\end{align*}
It follows that
\begin{align*}
\mu(z) & \le2^{-m}\cdot\left[\left(m^{t}-1\right)+1\right]\\
 & =2^{-\left(m-t\cdot\log m\right)}.
\end{align*}
Thus, $H_{\infty}(Z)\ge m-t\cdot\log m$ as required. Note that this
bound is a bit stronger than claimed in the lemma: indeed, we only
need the ``$-1$'' term in the lemma in order to deal with the case
where~$m=1$.
\end{myproof}

\subsubsection{Coupling}

Let $\mu_{1},\mu_{2}$ be two distributions over sample spaces $\Omega_{1},\Omega_{2}$.
A \emph{coupling} \emph{of $\mu_{1}$ and $\mu_{2}$} is a distribution
$\nu$ over the sample space $\Omega_{1}\times\Omega_{2}$ whose marginal
over the first coordinate is~$\mu_{1}$ and whose marginal over the
second coordinate is~$\mu_{2}$. In the case where $\Omega_{1}=\Omega_{2}=\Omega$,
the following standard fact allows us to use couplings to study the
statistical distance between $\mu_{1}$ and~$\mu_{2}$.
\begin{fact}
\label{fac:coupling}Let $\mu_{1},\mu_{2}$ be two distributions over
a sample space $\Omega$. The statistical distance between $\mu_{1}$
and $\mu_{2}$ is equal to the minimum, over all couplings $\nu$
of~$\mu_{1}$ and~$\mu_{2}$, of
\[
\Pr_{(X,Y)\gets\nu}\left[X\ne Y\right].
\]
\end{fact}
\noindent In particular, we can upper bound the statistical distance
between $\mu_{1}$ and~$\mu_{2}$ by constructing a coupling~$\nu$
in which the probability that $X\ne Y$ is small.

\subsection{Prefix-free codes}

A set of strings $C\subseteq\B^{*}$ is called a \emph{prefix-free
code} if no string in~$C$ is a prefix of another string in~$C$.
Given a string $w\in\B^{*}$, we denote its length by~$\left|w\right|$.
We use the following simple corollary of Kraft's inequality.
\begin{fact}
\label{fac:simplified-kraft}Let $C\subseteq\B^{*}$ be a finite prefix-free
code, and let $W$ be a random string taking values from~$C$. Then,
there exists a string $w\in C$ such that $\Pr\left[W=w\right]\ge\frac{1}{2^{\left|w\right|}}$.
\end{fact}
For completeness, we provide the following simple proof of \ref{fac:simplified-kraft}
that does not rely on Kraft's inequality.
\begin{myproof}
Let $n$ be the maximal length of a string in~$C$, and let $W'$
be a random string in~$\B^{n}$ that is sampled according to the
following process: sample a string $w$ from~$W$, choose a uniformly
distributed string~$z\in\B^{n-\left|w\right|}$, and set $W'=w\circ z$
(where here $\circ$ denotes string concatenation).

By a simple averaging argument, there exists a string $w'\in\B^{n}$
such that $\Pr\left[W'=w'\right]\ge\frac{1}{2^{n}}$. Since $C$ is
a prefix-free code, there exists a unique prefix~$w$ of~$w'$ that
is in~$C$. The definition of~$W'$ implies that
\[
\Pr\left[W'=w'\right]=\Pr\left[W=w\right]\cdot\frac{1}{2^{n-\left|w\right|}},
\]
because the only way the string $w'$ could be sampled is by first
sampling~$w$ and then sampling $z$ to be the rest of~$w'$ (again,
since $C$ is a prefix-free code). Hence, it follows that
\begin{align*}
\Pr\left[W=w\right]\cdot\frac{1}{2^{n-\left|w\right|}} & \ge\frac{1}{2^{n}}\\
\Pr\left[W=w\right] & \ge\frac{1}{2^{\left|w\right|}},
\end{align*}
as required.
\end{myproof}

\subsection{Discrepancy}

We start by recalling the definition of discrepancy.
\begin{repdefinition}{\ref{def:discrepancy}}
Let $\D$ be a finite set, let $g:\D\times\D\to\B$ be a function,
and let $U,V$ be independent random variables that are uniformly
distributed over~$\D$. Given a combinatorial rectangle $R\subseteq\D\times\D$,
the \emph{discrepancy of~$g$ with respect to~$R$}, denoted $\disc_{R}(g)$,
is defined as follows:
\[
\disc_{R}(g)=\left|\Pr\left[g(U,V)=0\text{ and }(U,V)\in R\right]-\Pr\left[g(U,V)=1\text{ and }(U,V)\in R\right]\right|.
\]
The \emph{discrepancy of~$g$}, denoted $\disc(g)$, is defined as
the maximum of $\disc_{R}(g)$ over all combinatorial rectangles~$R\subseteq\D\times\D$.
\end{repdefinition}
Let $g:\D\times\D\to\B$ be a function with discrepancy at most~$\left|\D\right|^{-\eta}$.
Such functions~$g$ satisfy the following ``extractor-like'' property.
In what follows, the parameter $\lambda$ controls $\b\left(g(X,Y)\right)$.
\begin{lemma}
\label{lem:discrepancy-extractor}Let $X,Y$ be independent random
variables taking values in~$\D$ such that $\Hm(X)+\Hm(Y)\ge(2-\eta+\lambda)\cdot\log\left|\D\right|$.
Then,
\[
\b\left(g(X,Y)\right)\le\left|\D\right|^{-\lambda}.
\]
\end{lemma}
\begin{myproof}
By \facref{flat-distributions}, it suffices to consider the case
where $X$ and~$Y$ have flat distributions. Let $A,B\subseteq\D$
be the sets over which $X,Y$ are uniformly distributed, and denote
$R\eqdef A\times B$. By the assumption on the min-entropies of~$X$
and~$Y$, it holds that $\left|R\right|\ge\left|\D\right|^{2-\eta+\lambda}$.

Let $U,V$ be random variables that are uniformly distributed over~$\D$.
Then, $X$ and $Y$ are distributed like $U\texttt{\ensuremath{\mid}}U\in A$
and $V\texttt{\ensuremath{\mid}}V\in B$ respectively. It follows
that
\begin{align*}
\b\left(g(X,Y)\right) & =\left|\Pr\left[g(X,Y)=0\right]-\Pr\left[g(X,Y)=1\right]\right|\\
 & =\left|\Pr\left[g(U,V)=0\texttt{\ensuremath{\mid}}(U,V)\in R\right]-\Pr\left[g(U,V)=1\texttt{\ensuremath{\mid}}(U,V)\in R\right]\right|\\
 & =\frac{\disc_{R}(g)}{\Pr\left[(U,V)\in R\right]}\\
 & \le\frac{\left|\D\right|^{-\eta}}{\Pr\left[(U,V)\in R\right]}\\
 & =\frac{\left|\D\right|^{-\eta}}{\left|R\right|/\left|\D\right|^{2}}\\
 & \le\frac{\left|\D\right|^{-\eta}}{\left|\D\right|^{-\eta+\lambda}}\\
 & =\left|\D\right|^{-\lambda},
\end{align*}
as required.
\end{myproof}
Using \lemref{discrepancy-extractor}, we can obtain the following
sampling property, which says that with high probability $X$ takes
a value~$x$ for which $\b\left(g(x,Y)\right)$ is small. In what
follows, the parameter $\lambda$ controls $\b\left(g(x,Y)\right)$,
the parameter~$\gamma$ controls the error probability, and recall
that $\eta$ is the parameter that controls the discrepancy of~$g$
(i.e., $\disc(g)\le2^{-\eta\cdot b}$).
\begin{lemma}
\label{lem:discrepancy-sampling}Let $\gamma,\lambda>0$. Let $X,Y$
be independent random variables taking values in~$\D$ such that
\[
\Hm(X)+\Hm(Y)\ge(2-\eta+\gamma+\lambda)\cdot\log\left|\D\right|+1.
\]
Then, the probability that $X$ takes a value~$x\in\D$ such that
\[
\b\left(g(x,Y)\right)>\left|\D\right|^{-\lambda}
\]
is less than $\left|\D\right|^{-\gamma}$.
\end{lemma}
\begin{myproof}
For every $x\in\D$, denote
\[
p_{x}\eqdef\Pr\left[g(x,Y)=1\right].
\]
Using this notation, our goal is to prove that
\[
\Pr_{X}\left[\left|p_{X}-\frac{1}{2}\right|\le\frac{1}{2}\left|\D\right|^{-\lambda}\right]<\left|\D\right|^{-\gamma}.
\]
We will prove that the probability that $p_{X}>\frac{1}{2}+\frac{1}{2}\left|\D\right|^{-\lambda}$
is less than $\frac{1}{2}\cdot\left|\D\right|^{-\gamma}$, and a similar
proof can be used to show that the probability that $p_{X}<\frac{1}{2}-\frac{1}{2}\left|\D\right|^{-\lambda}$
is less than $\frac{1}{2}\cdot\left|\D\right|^{-\gamma}$. The required
result will then follow by the union bound.

Let $\cE\subseteq\D$ be the set of values $x$ such that $p_{x}>\frac{1}{2}+\frac{1}{2}\left|\D\right|^{-\lambda}$.
Suppose for the sake of contradiction that $\Pr\left[X\in\cE\right]\ge\frac{1}{2}\cdot\left|\D\right|^{-\gamma}$.
It clearly holds that
\begin{equation}
\Pr\left[g(X,Y)=1\texttt{\ensuremath{\mid}}X\in\cE\right]>\frac{1}{2}+\frac{1}{2}\left|\D\right|^{-\lambda}.\label{eq:bias-of-biased-values}
\end{equation}
On the other hand, it holds that 
\begin{align*}
\Hm(X\texttt{\ensuremath{\mid}}X\in\cE) & \ge\Hm(X)-\log\frac{1}{\Pr\left[X\in\cE\right]}\\
\text{(By assumption on \ensuremath{\cE} and Fact \ref{fac:min-entropy-conditioning})} & \ge\Hm(X)-\gamma\cdot\log\left|\D\right|-1
\end{align*}
This implies that
\begin{align*}
\Hm(X\texttt{\ensuremath{\mid}}X\in\cE)+\Hm(Y) & \ge\Hm(X)+\Hm(Y)-\gamma\cdot\log\left|\D\right|-1\\
 & \ge(2-\eta)\cdot\log\left|\D\right|+\lambda\cdot\log\left|\D\right|.
\end{align*}

By \lemref{discrepancy-extractor}, it follows that
\[
\Pr\left[g(X,Y)=1\texttt{\ensuremath{\mid}}X\in\cE\right]\le\frac{1}{2}+\frac{1}{2}\left|\D\right|^{-\lambda},
\]
which contradicts Inequality~\ref{eq:bias-of-biased-values}. We
reached a contradiction, and therefore the probability that $p_{X}>\frac{1}{2}+\frac{1}{2}\left|\D\right|^{-\lambda}$
is less than $\frac{1}{2}\cdot\left|\D\right|^{-\gamma}$, as required.
\end{myproof}
Recall that $g^{\oplus I}(x,y)$ the function from $\cX^{I}\times\cY^{I}$
to $\B$ that outputs the parity of the string $g^{I}(x,y)$. We would
like to prove results like \lemref[s]{discrepancy-extractor} and~\ref{lem:discrepancy-sampling}
for functions of the form~$g^{\oplus I}$. To this end, we use the
following XOR lemma for discrepancy due to~\cite{LSS08}.
\begin{theorem}[\cite{LSS08}]
\label{thm:discrepancy-XOR-lemma}Let $m\in\N$. Then,
\[
\left(\disc(g)\right)^{m}\le\disc(g^{\oplus m})\le\left(64\cdot\disc(g)\right)^{m}.
\]
\end{theorem}
By combining \thmref{discrepancy-XOR-lemma} with \lemref[s]{discrepancy-extractor}
and~\ref{lem:discrepancy-sampling}, we obtain the following results.
\begin{corollary}
\label{cor:discrepancy-XOR-extractor}Let $\lambda>0$, $n\in\N$
and $S\subseteq\left[n\right]$. Let $X,Y$ be independent random
variables taking values in~$\D^{S}$ such that 
\[
\Hm(X)+\Hm(Y)\ge(2+\frac{6}{\log\left|\D\right|}-\eta+\lambda)\cdot\left|S\right|\cdot\log\left|\D\right|.
\]
Then,
\[
\b\left(g^{\oplus S}(X,Y)\right)\le\left|\D\right|^{-\lambda\cdot\left|S\right|}.
\]
\end{corollary}
\begin{corollary}
\label{cor:discrepancy-XOR-sampling}Let $\gamma,\lambda>0$, $n\in\N$
and $S\subseteq\left[n\right]$. Let $X,Y$ be independent random
variables taking values in~$\D^{S}$ such that 
\[
\Hm(X)+\Hm(Y)\ge(2+\frac{7}{\log\left|\D\right|}-\eta+\gamma+\lambda)\cdot\left|S\right|\cdot\log\left|\D\right|.
\]
Then, the probability that $X$ takes a value~$x\in\D$ such that
\[
\b\left(g^{\oplus S}(x,Y)\right)>\left|\D\right|^{-\lambda\cdot\left|S\right|}
\]
is less than $\left|\D\right|^{-\gamma\cdot\left|I\right|}$.
\end{corollary}

\section{\label{sec:lifting-machinery}Lifting Machinery}

In this section, we set up the machinery we need to prove our main
theorem, restated next.
\begin{reptheorem}{\ref{thm:main}}
For every $\eta>0$ there exists $c=O(\frac{1}{\eta^{2}}\cdot\log\frac{1}{\eta})$
such that the following holds: Let $\cS$ be a search problem that
takes inputs from~$\B^{n}$, and let $g:\B^{b}\times\B^{b}\to\B$
be an arbitrary function such that $\disc(g)\le2^{-\eta\cdot b}$
and such that $b\ge c\cdot\log n$. Then 
\[
\dcc(\cS\circ g^{n})=\Omega\left(\ddt(\cS)\cdot b\right),
\]
and for every $\varepsilon>0$ it holds that
\[
\rcc_{\varepsilon}(\cS\circ g^{n})=\Omega\left(\left(\rdt_{\varepsilon'}(\cS)-O(1)\right)\cdot b\right),
\]
where $\varepsilon'=\varepsilon+2^{-\eta\cdot b/8}$.
\end{reptheorem}
\noindent For the rest of this paper, we fix $\eta>0$ and let $c\in\N$
be some sufficiently large parameter that will be determined later
such that $c=O(\frac{1}{\eta^{2}}\cdot\log\frac{1}{\eta})$. Let $n\in\N$,
and let $g:\B^{b}\times\B^{b}\to\B$ be a function such that $\disc(g)\le2^{-\eta\cdot b}$
and such that $b\ge c\cdot\log n$. Note that when~$n=1$, the theorem
holds trivially, so we may assume that~$n\ge2$. For convenience,
we denote $\D\eqdef\B^{b}$ and $G\eqdef g^{n}$. Throughout the rest
of this section, $X$ and~$Y$ will always denote random variables
whose domain is~$\D^{n}$.

As explained in \Subsecref{our-techniques}, our simulation argument
is based on the idea that as long as the protocol did not transmit
too much information about the inputs, their distribution is similar
to the uniform distribution. The following definition, due to~\cite{GLMWZ},
formalizes the notion that the protocol did not transmit too much
information about an input $X$.
\begin{definition}
Let $\delta_{X}>0$. We say that a random variable $X$ is \emph{$\delta_{X}$-dense}
if for every $I\subseteq\left[n\right]$ it holds that $H_{\infty}(X_{I})\ge\delta_{X}\cdot b\cdot\left|I\right|$.
\end{definition}
\noindent As explained there, whenever the protocol transmits too
much information about a bunch of blocks $(X_{I},Y_{I})$ (where $I\subseteq\left[n\right]$),
the simulation queries $z_{I}$ and conditions the distribution on
$g(X_{I},Y_{I})=z_{I}$. The following definitions provide a useful
way for implementing this argument: restrictions are used to keep
track of which bits of~$z$ have been queried so far, and the notion
of structured variables expresses the desired properties of the distribution
of the inputs.
\begin{definition}
A \emph{restriction~$\rho$} is a string in $\left\{ 0,1,*\right\} ^{n}$.
We say that a coordinate~$i\in\left[n\right]$ is \emph{free} in~$\rho$
if $\rho_{i}=*$, and otherwise we say that $i$~is \emph{fixed}.
Given a restriction $\rho\in\left\{ 0,1,*\right\} ^{n}$, we denote
by $\free(\rho)$ and $\fix(\rho)$ the set of free and fixed coordinates
of~$\rho$ respectively. We say that a string $z\in\B^{n}$ is \emph{consistent}
with $\rho$ if $z_{\fix(\rho)}=\rho_{\fix(\rho)}$.
\end{definition}
Intuitively, $\fix(\rho)$ represents the queries that have been made
so far, and $\free(\rho)$ represents the coordinates that have not
been queried yet.
\begin{definition}[following \cite{GPW17}]
Let $\rho\in\left\{ 0,1,*\right\} ^{n}$ be a restriction, let $\tau>0$,
and let $X,Y$ be independent random variables. We say that $X$~and~$Y$
are $(\rho,\tau)$\emph{-structured} if there exist $\delta_{X},\delta_{Y}>0$
such that $X_{\free(\rho)}$ and $Y_{\free(\rho)}$ are $\delta_{X}$-dense
and $\delta_{Y}$-dense respectively, $\delta_{X}+\delta_{Y}\ge\tau$,
and
\[
g^{\fix(\rho)}\left(X_{\fix(\rho)},Y_{\fix(\rho)}\right)=\rho_{\fix(\rho)}.
\]
\end{definition}
We can now state our version of the uniform marginals lemma of~\cite{GPW17},
which formalizes the idea that if $X$ and~$Y$ are structured then
their distribution is similar to the uniform distribution over~$G^{-1}(z)$.
In what follows, the parameter~$\gamma$ controls the statistical
distance from the uniform distribution, and recall that $\eta$ is
the parameter that controls the discrepancy of~$g$ (i.e., $\disc(g)\le2^{-\eta\cdot b}$).
\begin{lemma}[Uniform marginals lemma]
\label{lem:uniform-marginals}There exists a universal constant~$h$
such that the following holds: Let $\gamma>0$, let $\rho$ be a restriction,
and let $z\in\B^{n}$ be a string that is consistent with~$\rho$.
Let $X,Y$ be independent random variables that are uniformly distributed
over sets $\cX,\cY\subseteq\D^{n}$ respectively, and assume that
they are $(\rho,\tau)$-structured where 
\[
\tau\ge2+\frac{h}{c}-\eta+\gamma.
\]
 Let $(X',Y')$ be uniformly distributed over $G^{-1}(z)\cap(\cX\times\cY)$.
Then, $X$ and~$Y$ are $2^{-\gamma\cdot b}$-close to $X'$ and~$Y'$
respectively.
\end{lemma}
\begin{remark}
\label{rem:choice-of-h}Here, as well as in other claims in the paper,
we denote by~$h$ some constant that is large enough to make the
proofs go through, and does not depend on any other parameter. The
constant~$h$ can be calculated explicitly, and we only refrain from
doing so in order to streamline the presentation. In all the cases
where we use this constant, it can be chosen to be at most~$50$.
\end{remark}
We defer the proof of \lemref{uniform-marginals} to \Subsecref{uniform-marginals},
and move to discuss the next issue. Recall that in order for $X$
and~$Y$ to be structured, the random variables $X_{\free(\rho)}$
and $Y_{\free(\rho)}$ have to be dense. However, as the simulation
progresses and the protocol transmits information, this property may
be violated, and $X_{\free(\rho)}$ or $Y_{\free(\rho)}$ may cease
to be dense. In order to restore the density, we use the following
folklore fact.
\begin{proposition}
\label{pro:density-restoring-fixing} Let $X$ be a random variable,
let $\delta_{X}>0$, and let $I\subseteq\left[n\right]$ be a maximal
subset of coordinates such that $H_{\infty}(X_{I})<\delta_{X}\cdot b\cdot\left|I\right|$.
Let $x_{I}\in\D^{I}$ be a value such that 
\[
\Pr\left[X_{I}=x_{I}\right]>2^{-\delta_{X}\cdot b\cdot\left|I\right|}.
\]
Then, the random variable $X_{\left[n\right]-I}\texttt{\ensuremath{\mid}}X_{I}=x_{I}$
is $\delta_{X}$-dense.
\end{proposition}
\begin{myproof}
Assume for the sake of contradiction that $X_{\left[n\right]-I}\texttt{\ensuremath{\mid}}X_{I}=x_{I}$
is not $\delta_{X}$-dense. Then, there exists a non-empty set $J\subseteq\left[n\right]-I$
such that $H_{\infty}(X_{J}\texttt{\ensuremath{\mid}}X_{I}=x_{I})<\delta_{X}\cdot b\cdot\left|J\right|$.
In particular, there exists a value $x_{J}\in\D^{J}$ such that 
\[
\Pr\left[X_{J}=x_{J}\texttt{\ensuremath{\mid}}X_{I}=x_{I}\right]>2^{-\delta_{X}\cdot b\cdot\left|J\right|}.
\]
But this implies that 
\[
\Pr\left[X_{I}=x_{I}\mathrm{~and~}X_{J}=x_{J}\right]>2^{-\delta_{X}\cdot b\cdot\left|I\cup J\right|},
\]
which means that 
\[
H_{\infty}(X_{I\cup J})<\delta_{X}\cdot b\cdot\left|I\cup J\right|.
\]
However, this contradicts the maximality of $I$.
\end{myproof}
\proref{density-restoring-fixing} is useful in the deterministic
setting, since in this setting the simulation is free to condition
the distributions of $X,Y$ in any way that maintains their density.
However, in the randomized setting, the simulation is more restricted,
and cannot condition the inputs on events such as $X_{I}=x_{I}$ which
may have very low probability. In \cite{GPW17}, this issue was resolved
by observing that the probability space can be partitioned to disjoint
events of the form~$X_{I}=x_{I}$, and that the randomized simulation
can use such a partition to achieve the same effect of \proref{density-restoring-fixing}.
This leads to the following lemma, which we use as well.
\begin{lemma}[Density-restoring partition \cite{GPW17}]
\label{lem:density-restoring-partition}Let $X$ be a random variable,
let $\cX$ denote the support of~$X$, and let $\delta_{X}>0$. Then,
there exists a partition 
\[
\cX\eqdef\cX^{1}\cup\dots\cup\cX^{\ell}
\]
where every $\cX^{j}$ is associated with a set $I_{j}\subseteq\left[n\right]$
and a value $x_{j}\in\D^{I_{j}}$ such that:
\begin{itemize}
\item $X_{I_{j}}\texttt{\ensuremath{\mid}}X\in\cX^{j}$ is fixed to~$x_{j}$.
\item $X_{\left[n\right]-I_{j}}\texttt{\ensuremath{\mid}}X\in\cX^{j}$ is
$\delta_{X}$-dense.
\end{itemize}
Moreover, if we denote $p_{\ge j}\eqdef\Pr\left[X\in\cX^{j}\cup\ldots\cup\cX^{\ell}\right]$,
then it holds that

\[
H_{\infty}(X_{\left[n\right]-I_{j}}\texttt{\ensuremath{\mid}}X\in\cX^{j})\ge H_{\infty}(X)-\delta_{X}\cdot b\cdot\left|I_{j}\right|-\log\frac{1}{p_{\ge j}}.
\]
\end{lemma}
We turn to discuss the ``conditioning issue'' that was discussed
in \Subsecref{our-techniques} and its resolution: As mentioned above,
the simulation uses \proref{density-restoring-fixing} and \lemref{density-restoring-partition}
to restore the density of the inputs by conditioning some of the blocks.
Specifically, suppose, for example, that $X_{\free(\rho)}$ is no
longer dense. Then, the simulation chooses appropriate $I\subseteq\free(\rho)$
and $x_{I}\in\D^{I}$, and conditions $X$ on the event $X_{I}=x_{I}$.
At this point, in order to make~$X$ and~$Y$ structured again,
we need to remove $I$ from~$\free(\rho)$, so the simulation queries
the bits in~$z_{I}$, and update the restriction $\rho$ by setting
$\rho_{I}=z_{I}$. Now, we have to make sure that $g(X_{I},Y_{I})=z_{I}$.
To this end, the simulation conditions $Y$ on the event $g(x_{I},Y_{I})=z_{I}$.
However, the latter conditioning reveals information about~$Y$,
which may have two harmful effects:
\begin{itemize}
\item \textbf{Leaking:} As discussed in \Subsecref{our-techniques}, our
analysis of the query complexity assumes that the protocol transmits
at most $C$ bits of information. It is important not to reveal more
information than that, or otherwise our query complexity may increase
arbitrarily. On average, we expect that conditioning on the event
$g(x_{I},Y_{I})=z_{I}$ would reveal only $\left|I\right|$~bits
of information, which is sufficiently small for our purposes. However,
there could be values of $x_{I}$ and $z_{I}$ for which much more
information is leaked. In this case, we say the conditioning is \emph{leaking}.
\item \textbf{Sparsifying:} Even if the conditioning reveals only $\left|I\right|$
bits of information on~$Y$, this could still ruin the density of~$Y$
if the set~$I$ is large. In this case, we say that the conditioning
is \emph{sparsifying}.
\end{itemize}
This is the ``conditioning issue'', and dealing with it is the technical
core of the paper. As explained in \Subsecref{our-techniques}, the
simulation deals with this issue by recognizing in advance which values
of~$X$ are ``dangerous'', in the sense that they may lead to a
bad conditioning, and discarding them before such conditioning may
take place. The foregoing discussion leads to the following definition
of a dangerous value.
\begin{definition}
\label{def:dangerous-value}Let $Y$ be a random variable taking values
from~$\D^{n}$. We say that a value~$x\in\Lambda^{n}$ is \emph{leaking}
if there exists a set $I\subseteq\left[n\right]$ and an assignment
$z_{I}\in\B^{I}$ such that 
\[
\Pr\left[g^{I}(x_{I},Y_{I})=z_{I}\right]<2^{-\left|I\right|-1}.
\]
Let $\delta_{Y},\varepsilon>0$, and suppose that $Y$ is $\delta_{Y}$-dense.
We say that a value $x\in\D^{n}$ is \emph{$\varepsilon$-sparsifying}
if there exists a set $I\subseteq\left[n\right]$ and an assignment
$z_{I}\in\B^{I}$ such that the random variable
\[
Y_{\left[n\right]-I}\texttt{\ensuremath{\mid}}g^{I}(x_{I},Y_{I})=z_{I}
\]
is not $(\delta_{Y}-\varepsilon)$-dense. We say that a value $x\in\D^{n}$
is $\varepsilon$\emph{-dangerous} if it is either leaking or $\varepsilon$-sparsifying.
\end{definition}
We can now state our main technical lemma, which says that $X$ has
only a small probability to take a dangerous value. This allows the
simulation to discard such values and resolve the conditioning issue.
In what follows, the parameter~$\gamma$ controls the error probability,
and recall that $\eta$ is the parameter that controls the discrepancy
of~$g$ (i.e., $\disc(g)\le2^{-\eta\cdot b}$).
\begin{lemma}[Main lemma]
\label{lem:main}There exists a universal constant\footnote{See \remref{choice-of-h} for further explanation on the constant~$h$.}~$h$
such that the following holds: Let $0<\gamma,\varepsilon,\tau\le1$
be such that $\tau\ge2+\frac{h}{c\cdot\varepsilon}-\eta+\gamma$ and
$\varepsilon\ge\frac{4}{b}$, and let $X,Y$ be $(\rho,\tau)$-structured
random variables. Then, the probability that $X_{\free(\rho)}$ takes
a value that is $\varepsilon$-dangerous for~$Y_{\free(\rho)}$ is
at most~$2^{-\gamma\cdot b}$.
\end{lemma}

\subsection{\label{subsec:uniform-marginals}Proof of the uniform marginals lemma}

Recall that the random variables $X$~and~$Y$ are $(\rho,\tau)$\emph{-structured}
if there exist $\delta_{X},\delta_{Y}>0$ such that $X_{\free(\rho)}$
and $Y_{\free(\rho)}$ are $\delta_{X}$-dense and $\delta_{Y}$-dense
respectively, $\delta_{X}+\delta_{Y}\ge\tau$, and $g^{\fix(\rho)}\left(X_{\fix(\rho)},Y_{\fix(\rho)}\right)=\rho_{\fix(\rho)}$.
In this section we prove the uniform marginals lemma, restated next.
\begin{replemma}{\ref{lem:uniform-marginals}}
There exists a universal constant~$h$ such that the following holds:
Let $\gamma>0$, let $\rho$ be a restriction, and let $z\in\B^{n}$
be a string that is consistent with~$\rho$. Let $X,Y$ be independent
random variables that are uniformly distributed over sets $\cX,\cY\subseteq\D^{n}$
respectively, and assume that they are $(\rho,\tau)$-structured where
\[
\tau\ge2+\frac{h}{c}-\eta+\gamma.
\]
 Let $(X',Y')$ be uniformly distributed over $G^{-1}(z)\cap(\cX\times\cY)$.
Then, $X$ and~$Y$ are $2^{-\gamma\cdot b}$-close to $X'$ and~$Y'$
respectively.
\end{replemma}
In order to prove \lemref{uniform-marginals}, we first prove the
following proposition, which says that the string $g^{\free(\rho)}(X_{\free(\rho)},Y_{\free(\rho)})$
is close to the uniform distribution in a very strong sense. In what
follows, the parameter~$\gamma$ controls the distance from the uniform
distribution, and recall that $\eta$ is the parameter that controls
the discrepancy of~$g$ (i.e., $\disc(g)\le2^{-\eta\cdot b}$).
\begin{proposition}[{Generalization of \cite[Lemma 13]{GLMWZ}}]
\label{pro:multiplicative-uniformity}There exists a universal constant~$h$
such that the following holds: Let $\gamma>0$. Let $X,Y$ be random
variables that are $(\rho,\tau)$-structured for $\tau\ge2+\frac{h}{c}-\eta+\gamma$,
and let $I\eqdef\free(\rho)$. Then, for every $z_{I}\in\B^{I}$ it
holds that 
\[
\Pr\left[g^{I}(X_{I},Y_{I})=z_{I}\right]\in(1\pm2^{-\gamma\cdot b})\cdot2^{-\left|I\right|}.
\]
\end{proposition}
\begin{myproof}
Let $h\eqdef8$. We use \corref{discrepancy-XOR-extractor} to upper
bound the biases of $g^{I}(X_{I},Y_{I})$, and then apply Vazirani's
lemma to show that it is close to the uniform distribution. Let $S\subseteq I$.
By assumption, the variables $X_{I},Y_{I}$ are $\delta_{X}$-dense
and $\delta_{Y}$-dense for some $\delta_{X},\delta_{Y}$ for which
$\delta_{X}+\delta_{Y}\ge2+\frac{8}{c}-\eta+\gamma$. Therefore, it
holds that
\[
\Hm(X_{S})+\Hm(Y_{S})\ge(2+\frac{6}{b}-\eta+\gamma+\frac{2}{c})\cdot b\cdot\left|S\right|
\]
and \corref{discrepancy-XOR-extractor} implies (with $\gamma=\gamma+\frac{2}{c}$)
that
\[
\b\left(g^{\oplus S}(X_{S},Y_{S})\right)\le2^{-\left(\gamma+\frac{2}{c}\right)\cdot b\cdot\left|S\right|}\le2^{-\gamma\cdot b}\cdot n^{-2\cdot\left|S\right|}\le2^{-\gamma\cdot b}\cdot(2\cdot\left|I\right|)^{-\left|S\right|}.
\]
Since the latter inequality holds for every $S\subseteq I$, it follows
by \lemref{vazirani} that
\[
\Pr\left[g^{I}(X_{I},Y_{I})=z_{I}\right]\in(1\pm2^{-\gamma\cdot b})\cdot2^{-\left|I\right|}
\]
for every $z_{I}\in\B^{I}$, as required.
\end{myproof}
We turn to prove the uniform marginals lemma.
\begin{myproof}[Proof of \lemref{uniform-marginals}]
Let $h'$ be the universal constant of \proref{multiplicative-uniformity}
and let $h\eqdef h'+2$. Let $(X',Y')$ be uniformly distributed over~$G^{-1}(z)\cap(\cX\times\cY)$,
and let $I\eqdef\free(\rho)$. We prove that $X$~is $2^{-\gamma\cdot b}$-close
to $X'$, and a similar argument works for~$Y$. Let $\cE\subseteq\cX$
be any test event. We show that
\[
\left|\Pr\left[X'\in\cE\right]-\Pr\left[X\in\cE\right]\right|\le2^{-\gamma\cdot b}.
\]
Without loss of generality we may assume that $\Pr\left[X\in\cE\right]\ge\frac{1}{2}$,
since otherwise we can replace~$\cE$ with its complement. Since
$X$ and~$Y$ are $(\rho,\tau)$-structured where 
\[
\tau\ge2+\frac{h}{c}-\eta+\gamma\ge2+\frac{h'}{c}-\eta+\gamma+\frac{2}{b},
\]
\proref{multiplicative-uniformity} implies that
\[
\Pr\left[g^{I}(X_{I},Y_{I})=z_{I}\right]\in(1\pm2^{-\gamma\cdot b-2})\cdot2^{-\left|I\right|}.
\]
Moreover, since $\Pr\left[X\in\cE\right]\ge\frac{1}{2}$, conditioning
on~$\cE$ cannot decrease the density of~$X$ by more than~$\frac{1}{b}$
(since this conditioning increases any probability by a factor of
at most~$2$). Therefore $X\texttt{\ensuremath{\mid}}\cE$ and~$Y$
together are $(\rho,\tau-\frac{1}{b})$-structured, where 
\[
\tau-\frac{1}{b}\ge2+\frac{h'}{c}-\eta+\gamma+\frac{1}{b}.
\]
 Hence, \proref{multiplicative-uniformity} implies that
\[
\Pr\left[g^{I}(X_{I},Y_{I})=z_{I}\texttt{\ensuremath{\mid}}X\in\cE\right]\in(1\pm2^{-\gamma\cdot b-1})\cdot2^{-\left|I\right|}.
\]
Now, it holds that
\begin{align*}
\Pr\left[X'\in\cE\right] & =\Pr\left[X\in\cE\texttt{\ensuremath{\mid}}G(X,Y)=z\right]\\
\text{(Bayes' formula)} & =\frac{\Pr\left[G(X,Y)=z\texttt{\ensuremath{\mid}}X\in\cE\right]}{\Pr\left[G(X,Y)=z\right]}\cdot\Pr\left[X\in\cE\right]\\
\text{(Since \ensuremath{g^{\fix(\rho)}(X_{\fix(\rho)},Y_{\fix(\rho)})=z_{\fix(\rho)}} by assumption)} & =\frac{\Pr\left[g^{I}(X_{I},Y_{I})=z_{I}\texttt{\ensuremath{\mid}}X\in\cE\right]}{\Pr\left[g^{I}(X_{I},Y_{I})=z_{I}\right]}\cdot\Pr\left[X\in\cE\right]\\
 & \le\frac{\left(1+2^{-\gamma\cdot b-2}\right)\cdot2^{-\left|I\right|}}{\left(1-2^{-\gamma\cdot b-1}\right)\cdot2^{-\left|I\right|}}\cdot\Pr\left[X\in\cE\right]\\
\text{(Since \ensuremath{\frac{1}{1-\alpha}\le1+2\alpha} for every \ensuremath{0<\alpha\le\frac{1}{2}})} & \le\left(1+2^{-\gamma\cdot b-2}\right)\cdot\left(1+2\cdot2^{-\gamma\cdot b-1}\right)\cdot\Pr\left[X\in\cE\right]\\
 & \le\Pr\left[X\in\cE\right]+2^{-\gamma\cdot b}.
\end{align*}
A similar calculation shows that 
\begin{align*}
\Pr\left[X'\in\cE\right] & \ge\frac{\left(1-2^{-\gamma\cdot b-2}\right)}{\left(1+2^{-\gamma\cdot b-1}\right)}\cdot\Pr\left[X\in\cE\right]\\
 & \ge\Pr\left[X\in\cE\right]-2^{-\gamma\cdot b}.
\end{align*}
It follows that
\[
\left|\Pr\left[X'\in\cE\right]-\Pr\left[X\in\cE\right]\right|\le2^{-\gamma\cdot b},
\]
as required.
\end{myproof}

\subsection{\label{subsec:main-technical-lemma}Proof of the main technical lemma}

In this section we prove our main technical lemma, which upper bounds
the probability of a variable to take a dangerous value. We first
recall the definition of a dangerous value and the lemma.
\begin{repdefinition}{\ref{def:dangerous-value}}
Let $Y$ be a random variable taking values from~$\D^{n}$. We say
that a value~$x\in\Lambda^{n}$ is \emph{leaking} if there exists
a set $I\subseteq\left[n\right]$ and an assignment $z_{I}\in\B^{I}$
such that 
\[
\Pr\left[g^{I}(x_{I},Y_{I})=z_{I}\right]<2^{-\left|I\right|-1}.
\]
Let $\delta_{Y},\varepsilon>0$, and suppose that $Y$ is $\delta_{Y}$-dense.
We say that a value $x\in\D^{n}$ is \emph{$\varepsilon$-sparsifying}
if there exists a set $I\subseteq\left[n\right]$ and an assignment
$z_{I}\in\B^{I}$ such that the random variable
\[
Y_{\left[n\right]-I}\texttt{\ensuremath{\mid}}g^{I}(x_{I},Y_{I})=z_{I}
\]
is not $(\delta_{Y}-\varepsilon)$-dense. We say that a value $x\in\D^{n}$
is $\varepsilon$\emph{-dangerous} if it is either leaking or $\varepsilon$-sparsifying.
\end{repdefinition}
\begin{replemma}{\ref{lem:main}}
There exists a universal constant~$h$ such that the following holds:
Let $0<\gamma,\varepsilon,\tau\le1$ be such that $\tau\ge2+\frac{h}{c\cdot\varepsilon}-\eta+\gamma$
and $\varepsilon\ge\frac{4}{b}$, and let $X,Y$ be $(\rho,\tau)$-structured
random variables. Then, the probability that $X_{\free(\rho)}$ takes
a value that is $\varepsilon$-dangerous for~$Y_{\free(\rho)}$ is
at most~$2^{-\gamma\cdot b}$.
\end{replemma}
Let $h$ be a universal constant that will be chosen to be sufficiently
large to make the inequalities in the proof hold. Let $\gamma,\varepsilon,\tau,\rho$
be as in the lemma, and assume that $X,Y$ are $(\rho,\tau)$-structured.
For simplicity of the presentation, we assume that all the coordinates
of~$\rho$ are free --- this can be assumed without loss of generality
since the fixed coordinates of~$\rho$ do not play any part in the
lemma. Thus, our goal is to prove an upper bound on the probability
that $X$ takes a value that is dangerous for~$Y$. By assumption,
there exist some parameters $\delta_{X},\delta_{Y}>0$ such that $X$
and~$Y$ are $\delta_{X}$-dense and $\delta_{Y}$-dense respectively,
and such that $\delta_{X}+\delta_{Y}\ge2+\frac{h}{c\cdot\varepsilon}-\eta-\gamma$.

We start by discussing the high-level ideas that underlie the proof.
We would like to prove an upper bound on the probability that $X$
takes a value that is either leaking or sparsifying. Proving the upper
bound for leaking values is relatively easy and is similar to the
proof of \proref{multiplicative-uniformity}: basically, since $X_{I}$
and~$Y_{I}$ are sufficiently dense, the string $g^{I}(X_{I},Y_{I})$
is multiplicatively close to uniform, which implies that most values
$x_{I}$ are non-leaking.

The more difficult task is to prove the upper bound for sparsifying
values. Basically, a value~$x$ is sparsifying if for some disjoint
$I,J\subseteq\free(\rho)$, conditioning on the value of $g^{I}(x_{I},Y_{I})$
decreases the min-entropy of $Y_{J}$ by more than $\varepsilon\cdot b\cdot\left|J\right|$
bits. Our first step is to apply Bayes' formula to the latter condition,
thus obtaining a more convenient condition to which we refer as ``skewing'':
a value~$x$ is skewing if conditioning on the value of~$Y_{J}$
decreases the min-entropy of $g^{I}(x_{I},Y_{I})$ by more than $\varepsilon\cdot b\cdot\left|J\right|$
bits --- in other words, the min-entropy of $g^{I}(x_{I},Y_{I})$
conditioned on $Y_{J}$ should be less than $\left|I\right|-\varepsilon\cdot b\cdot\left|J\right|$
(roughly).

It remains to prove an upper bound on the probability that $X$ takes
a skewing value. This requires proving a lower bound of roughly $\left|I\right|-\varepsilon\cdot b\cdot\left|J\right|$
on the min-entropy of $g^{I}(x_{I},Y_{I})\texttt{\ensuremath{\mid}}Y_{J}$
for most~$x$'s. By the min-entropy version of Vazirani's lemma (\lemref{vazirani-min-entropy}),
in order to prove this lower bound, it suffices to prove an upper
bound on the bias of $g^{S}(x_{S},Y_{S})\texttt{\ensuremath{\mid}}Y_{J}$
for every set $S\subseteq I$ for which\footnote{Recall that $c$ is a large constant such that $b\ge c\cdot\log n$.}
$\left|S\right|\gtrapprox\varepsilon\cdot c\cdot\left|J\right|$.

To this end, we use the ``extractor-like'' property of~$g^{S}$:
recall that by the discrepancy of~$g$ (\corref{discrepancy-XOR-sampling}),
the bias of $g^{S}(x_{S},Y_{S})\texttt{\ensuremath{\mid}}Y_{J}$ is
small for most~$x$'s whenever the min-entropy of $X_{S}$ and $Y_{S}\texttt{\ensuremath{\mid}}Y_{J}$
is high. Furthermore, recall that the min-entropy of $X_{S}$ and~$Y_{S}$
is high since we assumed that $X$ and~$Y$ are dense. The key step
is to observe that the min-entropy of $Y_{S}\texttt{\ensuremath{\mid}}Y_{J}$
is still high, since $S$ is large compared to~$J$. Thus, the min-entropy
of $X_{S}$ and $Y_{S}\texttt{\ensuremath{\mid}}Y_{J}$ is high, so
the bias of $g^{S}(x_{S},Y_{S})\texttt{\ensuremath{\mid}}Y_{J}$ is
small, and this implies the desired lower on the min-entropy of $g^{I}(x_{I},Y_{I})\texttt{\ensuremath{\mid}}Y_{J}$.

The argument we explained above almost works, except for a small issue:
We said that $\Hm(Y_{S}\texttt{\ensuremath{\mid}}Y_{J})$ is still
high, since $S$ is large compared to~$J$. Here, we implicitly assumed
that conditioning on the value of~$Y_{J}$ decreases the min-entropy
of~$Y_{S}$ by roughly $\left|J\right|\cdot b$~bits. This assumption
is true for the average value of~$Y_{J}$, but may fail for values
of~$Y_{J}$ that have a very small probability. In order to deal
with such values, we define a parameter $e_{y_{J}}$ which measures
the ``excess entropy'' of $y_{J}$, and keep track of it throughout
the proof. The key observation is that if we consider a value $y_{J}$
that has a small probability, then the criterion of ``skewing''
actually requires the min-entropy of $g^{I}(x_{I},Y_{I})$ to decrease
by roughly $\varepsilon\cdot b\cdot\left|J\right|+e_{y_{J}}$. Intuitively,
this means that the smaller the probability of $y_{J}$, the harder
it becomes for~$x$ to be skewing. After propagating the additional
term of~$e_{y_{J}}$ throughout our proof, we get that the set~$S$
can be assumed to satisfy 
\[
\left|S\right|\gtrapprox\varepsilon\cdot c\cdot\left|J\right|+\frac{e_{y_{J}}}{\log n}.
\]
This makes the set $S$ sufficiently large compared to $e_{y_{J}}$
that we can still deduce that $Y_{S}\texttt{\ensuremath{\mid}}Y_{J}=y_{J}$
has high min-entropy, which finished the argument. We now turn to
provide the formal proof, starting with a formal definition of the
parameter~$e_{y_{J}}$ and the criterion of ``skewing''.
\begin{definition}
\label{def:skewing}Recall that since $Y$ is $\delta_{Y}$-dense,
it holds that $\Pr\left[Y_{J}=y_{J}\right]\le2^{-\delta_{Y}\cdot b\cdot\left|J\right|}$
for every $J\subseteq\left[n\right]$ and $y_{J}\in\D^{J}$. We denote
by $e_{y_{J}}\in\R$ the (non-negative) number that satisfies
\[
\Pr\left[Y_{J}=y_{J}\right]=2^{-\delta_{Y}\cdot b\cdot\left|J\right|-e_{y_{J}}}.
\]
We say that a value $x\in\D^{n}$ is $\varepsilon$-skewing if there
exist disjoint non-empty sets $I,J\subseteq\left[n\right]$ and a
value $y_{J}\in\D^{J}$ such that
\[
H_{\infty}\left(\left.g^{I}(x_{I},Y_{I})\right|Y_{J}=y_{J}\right)<\left|I\right|-\varepsilon\cdot b\cdot\left|J\right|-e_{y_{J}}+1.
\]
\end{definition}
Next, we show that every dangerous value must be either leaking or
skewing by applying Bayes' formula.
\begin{claim}
\label{cla:skewing-condition}Let $x\in\D^{n}$ be an $\varepsilon$-dangerous
value that is not leaking for~$Y$. Then $x$ is $\varepsilon$-skewing.
\end{claim}
\begin{myproof}
Suppose that $x$ is $\varepsilon$-dangerous for~$Y$ and that it
is not leaking. We prove that $x$ is $\varepsilon$-skewing. By our
assumption, $x$ must be $\varepsilon$-sparsifying, so there exists
a set $I\subseteq\left[n\right]$ and an assignment $z_{I}\in\B^{I}$
such that the random variable
\[
Y_{\left[n\right]-I}\texttt{\ensuremath{\mid}}g^{I}(x_{I},Y_{I})=z_{I}
\]
is not $(\delta_{Y}-\varepsilon)$-dense. Thus, there exists a set
$J\subseteq\left[n\right]-I$ and a value $y_{J}\in\D^{n}$ such that
\[
\Pr\left[Y_{J}=y_{J}\texttt{\ensuremath{\mid}}g^{I}(x_{I},Y_{I})=z_{I}\right]>2^{-(\delta_{Y}-\varepsilon)\cdot b\cdot\left|J\right|}.
\]
By Bayes' formula, it holds that
\begin{align*}
\Pr\left[Y_{J}=y_{J}\texttt{\ensuremath{\mid}}g(x_{I},Y_{I})=z_{I}\right] & =\frac{\Pr\left[g^{I}(x_{I},Y_{I})=z_{I}\texttt{\ensuremath{\mid}}Y_{J}=y_{J}\right]\cdot\Pr\left[Y_{J}=y_{J}\right]}{\Pr\left[g(x_{I},Y_{I})=z_{I}\right]}\\
\text{(Definition of \ensuremath{e_{y_{J}}})} & =\frac{\Pr\left[g^{I}(x_{I},Y_{I})=z_{I}\texttt{\ensuremath{\mid}}Y_{J}=y_{J}\right]\cdot2^{-\delta_{Y}\cdot b\cdot\left|J\right|-e_{y_{J}}}.}{\Pr\left[g(x_{I},Y_{I})=z_{I}\right]}\\
\text{(Since \ensuremath{x} is not leaking for \ensuremath{Y})} & \le\frac{\Pr\left[g^{I}(x_{I},Y_{I})=z_{I}\texttt{\ensuremath{\mid}}Y_{J}=y_{J}\right]\cdot2^{-\delta_{Y}\cdot b\cdot\left|J\right|-e_{y_{J}}}.}{2^{-\left|I\right|-1}}.
\end{align*}
Hence, it follows that
\[
\frac{\Pr\left[g^{I}(x_{I},Y_{I})=z_{I}\texttt{\ensuremath{\mid}}Y_{J}=y_{J}\right]\cdot2^{-\delta_{Y}\cdot b\cdot\left|J\right|-e_{y_{J}}}}{2^{-\left|I\right|-1}}>2^{-(\delta_{Y}-\varepsilon)\cdot b\cdot\left|J\right|},
\]
which implies that
\[
\Pr\left[g^{I}(x_{I},Y_{I})=z_{I}\texttt{\ensuremath{\mid}}Y_{J}=y_{J}\right]>2^{-\left|I\right|+\varepsilon\cdot b\cdot\left|J\right|+e_{y_{J}}-1}.
\]
This means that
\[
H_{\infty}\left(\left.g^{I}(x_{I},Y_{I})\right|Y_{J}=y_{J}\right)<\left|I\right|-\varepsilon\cdot b\cdot\left|J\right|-e_{y_{J}}+1.
\]
That is, $x$~is $\varepsilon$-skewing, as required.
\end{myproof}
As explained above, we will upper bound the probability of dangerous
values by upper bounding the biases of~$g(x_{S},Y_{S})\texttt{\ensuremath{\mid}}Y_{J}$
for every $S\subseteq I$. To this end, it is convenient to define
the notion of a ``biasing value'', which is a value~$x$ for which
one of the biases is too large.
\begin{definition}
\label{def:biasing}We say that a value $x\in\D^{n}$ is \emph{biasing
(for~$Y$) with respect to }disjoint sets $S,J\subseteq\left[n\right]$
and an assignment $y_{J}\in\D^{J}$ if
\[
\b\left(g^{\oplus S}(x_{S},Y_{S})\texttt{\ensuremath{\mid}}Y_{J}=y_{J}\right)>\frac{1}{2}\cdot\left(2n\right)^{-\left|S\right|}.
\]
We say that $x$ is \emph{$\varepsilon$-biasing (for~$Y$) with
respect to} a set $S\subseteq\left[n\right]$ if there exists a set
$J\subseteq\left[n\right]-S$ and an assignment $y_{J}\in\D^{J}$
that satisfy
\begin{equation}
\left|S\right|\ge c\cdot\varepsilon\cdot\left|J\right|+\frac{e_{y_{J}}+2}{\log n}\label{eq:skewing-size-bound}
\end{equation}
such that $x$ is biasing with respect to $S$, $J$, and $y_{J}$
(if $J$ is the empty set, we define $e_{y_{J}}=0$). Finally, we
say that $x$ is \emph{$\varepsilon$-biasing} (for~$Y$) if there
exists a non-empty set $S$ with respect to which $x$~is $\varepsilon$-biasing.
\end{definition}
We now apply the min-entropy version of Vazirani's lemma to show that
values that are not biasing are not dangerous.
\begin{claim}
\label{cla:biasing-condition}If a value $x\in\D^{n}$ is not $\varepsilon$-biasing
for~$Y$ then it is not $\varepsilon$-dangerous for~$Y$.
\end{claim}
\begin{myproof}
Suppose that $x\in\D^{n}$ is a value that is not $\varepsilon$-biasing
for~$Y$. We prove that $x$ is not $\varepsilon$-dangerous for~$Y$.
We start by proving that $x$ is not leaking. Let $I\subseteq\left[n\right]$
and let $z_{I}\in\B^{I}$. We wish to prove that
\[
\Pr\left[\Pr\left[g^{I}(x_{I},Y_{I})=z_{I}\right]\ge2^{-\left|I\right|-1}\right].
\]
Observe that, by the assumption that $x$ is not $\varepsilon$-biasing,
it holds for every non-empty set $S\subseteq I$ that 
\[
\b\left(g^{\oplus S}(x_{S},Y_{S})\right)\le\frac{1}{2}\cdot\left(2n\right)^{-\left|S\right|}
\]
(this follows by substituting $J=\emptyset$ in the definition of
$\varepsilon$-biasing and noting that in this case $e_{y_{J}}=0$).
It now follows from \lemref{vazirani-min-entropy} that $\Pr\left[g^{I}(x_{I},Y_{I})=z_{I}\right]\ge2^{-\left|I\right|-1}$,
as required.

We turn to prove that $x$ is not $\varepsilon$-skewing. Let $I,J\subseteq\left[n\right]$
be disjoint sets and let $y_{J}\in\D^{J}$ be an assignment. We wish
to prove that
\[
H_{\infty}\left(\left.g^{I}(x_{I},Y_{I})\right|Y_{J}=y_{J}\right)\ge\left|I\right|-\varepsilon\cdot b\cdot\left|J\right|-e_{y_{J}}+1.
\]
By \lemref{vazirani-min-entropy}, it suffices to prove that for every
set $S\subseteq I$ such that $\left|S\right|\ge\frac{\varepsilon\cdot b\cdot\left|J\right|+e_{y_{J}}+2}{\log n}$
it holds that
\[
\b\left(g^{\oplus S}(x_{S},Y_{S})\texttt{\ensuremath{\mid}}Y_{J}=y_{J}\right)\le\left(2n\right)^{-\left|S\right|}.
\]
To this end, observe that every such set $S$ satisfies
\[
\left|S\right|\ge\varepsilon\cdot c\cdot\left|J\right|+\frac{e_{y_{J}}+2}{\log n},
\]
and since by assumption $x$ is not $\varepsilon$-biasing with respect
to $S$, the required upper bound on the bias must hold. It follows
that $x$ is neither leaking nor $\varepsilon$-skewing, and therefore
it is not $\varepsilon$-dangerous, as required.
\end{myproof}
Finally, we prove an upper bound on the probability of~$X$ to take
an $\varepsilon$-biasing value, which together with \claref{biasing-condition}
implies \lemref{main}. As explained above, the idea is to combine
the discrepancy of $g$ with the observation that $X_{S}$ and $Y_{S}$
have large min-entropy even conditioned on~$Y_{J}=y_{J}$ (which
holds since $X,Y$ are dense and $S$ is large compared to $\left|J\right|$
and $e_{y_{J}}$).
\begin{proposition}
The probability that $X$ takes a value~$x$ that is $\varepsilon$-biasing
for~$Y$ is at most~$2^{-\gamma\cdot b}$.
\end{proposition}
\begin{myproof}
We begin with upper bounding the probability of $X$ to take a value
that is $\varepsilon$-biasing \emph{with respect to specific choices
of $S$, $J$, and $y_{J}$}, and the rest of the proof will follow
by applying union bounds over all possible choices of $S$, $J$,
and $y_{J}$. Let $S,J\subseteq\left[n\right]$ be disjoint sets and
let $y_{J}\in\D^{J}$ be an assignment such that $S$, $J$, and $y_{J}$
together satisfy Equation~\ref{eq:skewing-size-bound}, i.e.,
\[
\left|S\right|\ge c\cdot\varepsilon\cdot\left|J\right|+\frac{e_{y_{J}}+2}{\log n},
\]
For simplicity, we assume that $J$ is non-empty (in the case where
$J$ is empty, the argument is similar but simpler). Since we assumed
that $\varepsilon\ge\frac{4}{b}$ and that $J$ is non-empty, and
it holds that $\frac{1}{2}\cdot c\cdot\varepsilon\cdot\left|J\right|\ge\frac{2}{\log n}$
and therefore
\[
\left|S\right|\ge\frac{1}{2}\cdot c\cdot\varepsilon\cdot\left|J\right|+\frac{e_{y_{J}}}{\log n}.
\]
In other words, it holds that
\begin{equation}
\left|S\right|\cdot\log n\ge\frac{1}{2}\cdot\varepsilon\cdot b\cdot\left|J\right|+e_{y_{J}}.\label{eq:skewing-size-bound-rephrased}
\end{equation}
By assumption, $Y$ is $\delta_{Y}$-dense, so $H_{\infty}(Y_{S})\ge\delta_{Y}\cdot b\cdot\left|S\right|$.
By \facref{min-entropy-conditioning}, it follows that
\begin{align*}
H_{\infty}(Y_{S}\texttt{\ensuremath{\mid}}Y_{J}=y_{J}) & \ge\delta_{Y}\cdot b\cdot\left|S\right|-\log\frac{1}{\Pr\left[Y_{J}=y_{J}\right]}\\
 & =\delta_{Y}\cdot b\cdot\left|S\right|-\left(\delta_{Y}\cdot b\cdot\left|J\right|+e_{y_{J}}\right)\\
 & \ge\delta_{Y}\cdot b\cdot\left|S\right|-\left(b\cdot\left|J\right|+e_{y_{J}}\right)\\
 & \ge\delta_{Y}\cdot b\cdot\left|S\right|-\frac{2}{\varepsilon}\cdot\left(\frac{1}{2}\cdot\varepsilon\cdot b\cdot\left|J\right|+e_{y_{J}}\right)\\
\text{(By Equation \ref{eq:skewing-size-bound}) } & \ge\delta_{Y}\cdot b\cdot\left|S\right|-\frac{2}{\varepsilon}\cdot\left|S\right|\cdot\log n\\
\text{(Since \ensuremath{b\ge c\cdot\log n})} & \ge\left(\delta_{Y}-\frac{2}{c\cdot\varepsilon}\right)\cdot b\cdot\left|S\right|
\end{align*}
Moreover, $X$ is $\delta_{X}$-dense and thus
\[
H_{\infty}(X_{S})+H_{\infty}(Y_{S}\texttt{\ensuremath{\mid}}Y_{J}=y_{J})\ge\left(\delta_{X}+\delta_{Y}-\frac{2}{c\cdot\varepsilon}\right)\cdot b\cdot\left|S\right|\ge\left(2+\frac{8}{c\cdot\varepsilon}-\eta+\gamma\right)\cdot b\cdot\left|S\right|,
\]
where the second inequality is made to hold for by choosing $h$ to
be sufficiently large. It follows by \corref{discrepancy-XOR-sampling}
(with $\lambda=\frac{3}{c\cdot\varepsilon}$ and $\gamma=\gamma+\frac{5}{c\cdot\varepsilon}$)
that the probability that $X_{S}$ takes a value $x_{S}\in\D^{S}$
for which
\begin{equation}
\b\left(g^{\oplus S}(\alpha,Y_{S})\texttt{\ensuremath{\mid}}Y_{J}=y_{J}\right)>\frac{1}{2}\cdot\left(2n\right)^{-\left|S\right|}\ge2^{-3\log n\cdot\left|S\right|}\ge2^{-\frac{3}{c\cdot\varepsilon}\cdot b\cdot\left|S\right|}\label{eq:skewing-biased}
\end{equation}
is at most $2^{-\left(\gamma+\frac{5}{c\cdot\varepsilon}\right)\cdot b\cdot\left|S\right|}$.

We turn to applying the union bounds. First, we show that for every
$S\subseteq\left[n\right]$, the probability that $X$ takes a value
that is $\varepsilon$-biasing with respect to $S$ is at most~$2^{-(\gamma+\frac{2}{c\cdot\varepsilon})\cdot b\cdot\left|S\right|}$
by taking upper bound over all choices of $J$ and $y_{J}$. Note
that we only need to consider sets $J\subseteq\left[n\right]$ for
which $\left|J\right|\le\frac{1}{c\cdot\varepsilon}\cdot\left|S\right|$.
It follows that the probability that $X_{S}$ takes a value that satisfies
Equation~\ref{eq:skewing-biased} for some $J$ and~$y_{J}$ is
at most
\begin{align*}
 & \sum_{J\subseteq\left[n\right]:\left|J\right|\le\frac{1}{c\cdot\varepsilon}\cdot\left|S\right|}\sum_{y_{J}\in\D^{J}}2^{-(\gamma+\frac{5}{c\cdot\varepsilon})\cdot b\cdot\left|S\right|}\\
 & \le\sum_{j=1}^{\frac{1}{c\cdot\varepsilon}\cdot\left|S\right|}\binom{n}{j}\cdot2^{b\cdot j}\cdot2^{-(\gamma+\frac{5}{c\cdot\varepsilon})\cdot b\cdot\left|S\right|+1}\\
 & \le n\cdot\binom{n}{\frac{1}{c\cdot\varepsilon}\cdot\left|S\right|}\cdot2^{\frac{1}{c\cdot\varepsilon}\cdot\left|S\right|\cdot b}\cdot2^{-(\gamma+\frac{5}{c\cdot\varepsilon})\cdot b\cdot\left|S\right|}\\
 & \le2^{-\left(\gamma+\frac{5}{c\cdot\varepsilon}\right)\cdot b\cdot\left|S\right|+\frac{1}{c\cdot\varepsilon}\cdot\left|S\right|\cdot\left(\log n+b\right)+\log n}\\
 & \le2^{-(\gamma+\frac{2}{c\cdot\varepsilon})\cdot b\cdot\left|S\right|},
\end{align*}
where the last inequality follows since
\[
\frac{1}{c\cdot\varepsilon}\cdot\left|S\right|\cdot\left(\log n+b\right)+\log n\le\frac{1}{c\cdot\varepsilon}\cdot2b\cdot\left|S\right|+b\le\frac{3}{c\cdot\varepsilon}\cdot b\cdot\left|S\right|.
\]
The above calculation showed that the probability that $X$ takes
a value that is $\varepsilon$-biasing with respect to a fixed set~$S$
is at most~$2^{-(\gamma+\frac{2}{c\cdot\varepsilon})\cdot b\cdot\left|S\right|}$.
Taking a union bound over all non-empty sets $S\subseteq\left[n\right]$,
the probability that $X$ takes a value that is $\varepsilon$-biasing
for~$Y$ is at most
\begin{align*}
\sum_{\emptyset\ne S\subseteq\left[n\right]}2^{-(\gamma+\frac{2}{c\cdot\varepsilon})\cdot b\cdot\left|S\right|} & \le\sum_{s=1}^{n}\binom{n}{s}\cdot2^{-(\gamma+\frac{2}{c\cdot\varepsilon})\cdot b\cdot s}\\
 & \le\sum_{s=1}^{n}2^{-(\gamma+\frac{2}{c\cdot\varepsilon})\cdot b\cdot s-s\cdot\log n}\\
\text{(\ensuremath{\frac{b}{c\cdot\varepsilon}\ge\log n})} & \le\sum_{s=1}^{n}2^{-(\gamma+\frac{1}{c\cdot\varepsilon})\cdot b\cdot s}\\
 & \le2^{-(\gamma+\frac{1}{c\cdot\varepsilon})\cdot b+1}\\
 & \le2^{-\gamma\cdot b}.
\end{align*}

We have thus shown that the probability that $X$ takes a value that
is $\varepsilon$-biasing is at most~$2^{-\gamma\cdot b}$, as required.
\end{myproof}

\section{\label{sec:deterministic-lifting}The deterministic lifting theorem}

In this section, we prove the deterministic part of our main theorem.
In fact, we prove the following more general result.
\begin{theorem}[Deterministic lifting theorem]
For every $\eta>0$ there exists $c=O(\frac{1}{\eta^{2}})$ such
that the following holds: Let $n\in\N$ be such that $n\ge2$, let
$\D\eqdef\B^{b}$ be such that $b\ge c\cdot\log n$, let $g:\D\times\D\to\B$
be a function such that $\disc(g)\le2^{-\eta\cdot b}$, and let $G=g^{n}$.
Let $\Pi$ be a deterministic protocol that takes inputs in $\D^{n}\times\D^{n}$
and that has communication complexity~$C$ and round complexity~$r$.
Then, there exists a deterministic parallel decision tree~$T$ that
that on input $z\in\B^{n}$ outputs a transcript~$\pi$ of~$\Pi$
that is consistent with some pair of inputs $(x,y)\in G^{-1}(z)$,
and that has query complexity $O(\frac{C}{b})$ and depth~$r$.
\end{theorem}
Observe that this theorem implies the lower bound of the main theorem:
Given a protocol~$\Pi$ that solves~$\cS\circ G$ with complexity~$C$,
we use the theorem to construct a tree~$T$ that on input~$z$ outputs
the output of~$\Pi$ on some pair of inputs in~$G^{-1}(z)$. This
tree~$T$ clearly solves $\cS$, and the query complexity of~$T$
is $O(\frac{C}{b})$. This implies that $\ddt(\cS)=O\left(\dcc(\cS\circ G)/b\right)$,
or in other words, $\dcc(\cS\circ G)=\Omega\left(\ddt(\cS)\cdot b\right)$,
as required.

For the rest of this section, fix $\Pi$ to be an arbitrary deterministic
protocol that takes inputs in $\D^{n}\times\D^{n}$, and denote by
$C$ and~$r$ its communication complexity and round complexity respectively.
The rest of this section is organized as follows: We first describe
the construction of the parallel decision tree~$T$ in \Subsecref{deterministic-construction}.
We then prove that the output of~$T$ is always correct in \Subsecref{deterministic-correctness}.
Finally, we upper bound the query complexity of~$T$ in \Subsecref{deterministic-complexity}.

\subsection{\label{subsec:deterministic-construction}The construction of~$T$}

Let $h'$ be the maximum among the universal constants of \proref{multiplicative-uniformity}
and the main technical lemma (\lemref{main}), and let $h$ be a universal
constant that will be chosen to be sufficiently large to make the
inequalities in the proof hold. Let $\varepsilon\eqdef\frac{h}{c\cdot\eta}$,
let $\delta\eqdef1-\frac{\eta}{4}+\frac{\varepsilon}{2}$, and let
$\tau\eqdef2\cdot\delta-\varepsilon$. The tree~$T$ constructs the
transcript~$\pi$ by simulating the protocol~$\Pi$ round-by-round,
each time adding a single message to~$\pi$. Throughout the simulation,
the tree maintains a rectangle~$\cX\times\cY\subseteq\D^{n}\times\D^{n}$
of inputs that are consistent with~$\pi$ (but not necessarily of
all such inputs). In what follows, we denote by $X$ and~$Y$ random
variables that are uniformly distributed over $\cX$ and~$\cY$ respectively.
The tree will maintain the invariant that $X$ and~$Y$ are $(\rho,\tau)$-structured,
where $\rho$ is a restriction that keeps track of the queries the
tree has made so far. In fact, the tree will maintain a more specific
invariant: whenever it is Alice's turn to speak, $X_{\free(\rho)}$
is $(\delta-\varepsilon)$-dense and $Y_{\free(\rho)}$ is $\delta$-dense,
and whenever it is Bob's turn to speak, the roles of $X$ and~$Y$
are reversed.

When the tree~$T$ starts the simulation, the tree sets the transcript
$\pi$ to be the empty string, the restriction $\rho$ to~$\left\{ *\right\} ^{n}$,
and the sets $\cX,\cY$ to~$\D^{n}$. At this point the invariant
clearly holds. We now explain how $T$ simulates a single round of
the protocol while maintaining the invariant. Suppose that the invariant
holds at the beginning of the current round, and assume without loss
of generality that it is Alice's turn to speak. The tree~$T$ performs
the following steps:
\begin{enumerate}
\item \label{enu:deterministic-simulation-dangerous-values}The tree conditions~$X_{\free(\rho)}$
on not taking a value that is $\varepsilon$-dangerous for $Y_{\free(\rho)}$
(i.e., the tree removes from~$\cX$ all the values $x$ for which
$x_{\free(\rho)}$ is $\varepsilon$-dangerous for~$Y_{\free(\rho)}$).
\item \label{enu:deterministic-simulation-choosing-message}The tree~$T$
chooses an arbitrary message~$M$ of Alice with the following property:
the probability of Alice sending $M$ on input~$X$ is at least $2^{-\left|M\right|}$
(the existence of~$M$ will be justified soon). The tree adds~$M$
to the transcript~$\pi$, and conditions $X$ on the event of sending~$M$
(i.e., the tree sets~$\cX$ to be the subset of inputs that are consistent
with~$M$).
\item \label{enu:deterministic-simulation-taking-heavy-value}Let $I\subseteq\free(\rho)$
be a maximal set that violates the $\delta$-density of~$X_{\free(\rho)}$
(i.e., $H_{\infty}(X_{I})<\delta\cdot b\cdot\left|I\right|$), and
let $x_{I}\in\D^{I}$ be a value that satisfies $\Pr\left[X_{I}=x_{I}\right]>2^{-\delta\cdot b\cdot\left|I\right|}$.
The tree conditions $X$~on $X_{I}=x_{I}$ (i.e., the tree removes
from~$\cX$ all the values that are inconsistent with that event).
By \proref{density-restoring-fixing}, $X_{\free(\rho)-I}$ is now
$\delta$-dense.
\item \label{enu:deterministic-simulation-query}The tree queries~$z_{I}$,
and updates~$\rho$ accordingly.
\item \label{enu:deterministic-simulation-consistent-with-queries}The tree
conditions $Y$~on $g^{I}(x_{I},Y_{I})=\rho_{I}$ (i.e., the tree
sets $\cY$ to be the subset of values~$y$ for which $g^{I}(x_{I},y_{I})=\rho_{I}$).
Due to Step~\enuref{deterministic-simulation-dangerous-values},
the variable~$X_{\free(\rho)}$ must take a value that is not $\varepsilon$-dangerous,
and therefore $Y_{\free(\rho)}$ is necessarily $(\delta-\varepsilon)$-dense.
\end{enumerate}
After those steps take place, it becomes Bob's turn to speak, and
indeed, $X_{\free(\rho)}$ and $Y_{\free(\rho)}$ are $\delta$-dense
and $(\delta-\varepsilon)$-dense respectively. Thus, the invariant
is maintained. When the protocol~$\Pi$ stops, the tree~$T$ outputs
the transcript~$\pi$ and halts. In order for the foregoing construction
to be well-defined, it remains to explain three points:
\begin{itemize}
\item First, we should explain why the set~$\cX$ remains non-empty after
Step~\enuref{deterministic-simulation-dangerous-values} (otherwise,
the following steps are not well-defined). To this end, recall that
$X$ and~$Y$ are $(\rho,\tau)$-structured and observe that $\tau$
can be made larger than $2+\frac{h'}{c\cdot\varepsilon}-\eta$ by
choosing $h$ to be sufficiently large (see \Subsecref{deterministic-complexity}
for a detailed calculation). Hence, by our main lemma (\lemref{main}),
the variable $X_{\free(\rho)}$ has a non-zero probability to take
a value that is not $\varepsilon$-dangerous for~$Y_{\free(\rho)}$,
so $\cX$ is non-empty after this step.
\item Second we should explain why the message~$M$ in Step~\enuref{deterministic-simulation-choosing-message}
exists. To see why, observe that the set of possible messages of Alice
forms a prefix-free code --- otherwise, Bob will not be able to tell
when Alice finished speaking and his turn starts. Hence, by \facref{simplified-kraft},
it follows that there exists a message~$M$ with probability at least~$2^{-\left|M\right|}$.
\item Third, we should explain why the set~$\cY$ remains non-empty after
Step~\enuref{deterministic-simulation-consistent-with-queries}.
To this end, recall that $X$ must take a value that is not $\varepsilon$-dangerous
for~$Y$, and in particular, the value of~$X$ is necessarily not
leaking. This means that in particular that the string $g^{I}(x_{I},Y_{I})$
has non-zero probability to be equal to $\rho_{I}$, so $\cY$ is
non-empty after this step.
\end{itemize}

\paragraph*{The depth of~$T$.}

We now observe that the depth of~$T$ is equal to the round complexity
of~$\Pi$. Note that in each round, the tree~$T$ issues a set of
queries~$I$ simultaneously. Thus, $T$ is a parallel decision tree
whose depth equals the maximal number of rounds of~$\Pi$, as required.

\subsection{\label{subsec:deterministic-correctness}The correctness of~$T$}

We now prove that when the decision tree~$T$ halts, the transcript~$\pi$
is consistent with some inputs $(x,y)\in G^{-1}(z)$. Clearly, the
transcript~$\pi$ is consistent with all the inputs in the rectangle~$\cX\times\cY$.
Thus, it suffices to show that there exist $x\in\cX$ and~$y\in\cY$
such that $G(x,y)=z$. To this end, recall that when the tree halts,
the random variables $X$ and~$Y$ are $(\rho,\tau)$-structured.
Since $\rho$ is consistent with~$z$, it holds for every $x\in\cX$
and $y\in\cY$ that
\begin{equation}
g^{\fix(\rho)}(x_{\fix(\rho)},y_{\fix(\rho)})=z_{\fix(\rho)}.\label{eq:deterministic-fixed-z}
\end{equation}
 It remains to deal with the free coordinates of~$\rho$. Since $\tau$
can be made larger than $2+\frac{h'}{c}-\eta$ by choosing $h$ to
be sufficiently large (see \subsecref{deterministic-complexity} for
a detailed calculation), it follows by \proref{multiplicative-uniformity}
that
\[
\Pr\left[g^{\free(\rho)}(x_{\free(\rho)},y_{\free(\rho)})=z_{\free(\rho)}\right]>0.
\]
In particular, there exist $x\in\cX$ and $y\in\cY$ such that
\begin{equation}
g^{\free(\rho)}(x_{\free(\rho)},y_{\free(\rho)})=z_{\free(\rho)}.\label{eq:deterministic-free-z}
\end{equation}
By combining Equations \eqref{deterministic-fixed-z} and~\eqref{deterministic-free-z},
we get that $G(x,y)=z$, as required.

\subsection{\label{subsec:deterministic-complexity}The query complexity of~$T$}

We conclude by showing that the total number of queries the tree~$T$
makes is at most $O(\frac{C}{b})$. To this end, we define the deficiency
of~$X,Y$ to be 
\[
2\cdot b\cdot\left|\free(\rho)\right|-H_{\infty}(X_{\free(\rho)})-H_{\infty}(Y_{\free(\rho)}).
\]
We will prove that whenever the protocol transmits a message~$M$,
the deficiency increases by~$O(\left|M\right|)$, and that whenever
the tree~$T$ makes a query, the deficiency is decreased by~$\Omega(b)$.
Since the deficiency is always non-negative, and the protocol transmits
at most $C$~bits, it will follow that the tree must make at most
$O(\frac{C}{b})$ queries. More specifically, we prove that in every
round, the first two steps from \subsecref{deterministic-construction}
increase the deficiency by at most~$\left|M\right|+1$ in total,
and the rest of the steps decrease the deficiency by at least~$\Omega(\left|I\right|\cdot b)$,
and this will imply the desired result.

Fix a round of the simulation, and assume without loss of generality
that the message is sent by Alice. We start by analyzing Step~\enuref{deterministic-simulation-dangerous-values}.
At this step, the tree conditions~$X_{\free(\rho)}$ on taking dangerous
values that are not $\varepsilon$-dangerous for~$Y_{\free(\rho)}$.
We show that this step increases the deficiency by at most one bit.
By applying our main technical lemma (\lemref{main}) with $\gamma=\frac{1}{b}$,
it follows that the probability that $X_{\free(\rho)}$ is $\varepsilon$-dangerous
is at most~$\frac{1}{2}$. By \facref{min-entropy-conditioning},
it follows that conditioning on non-dangerous values decreases $\Hm(X_{\free(\rho)})$
by at most one bit, and therefore it increases the deficiency by at
most one bit. To see why we can apply the main lemma with $\gamma=\frac{1}{b}$,
recall that at this point $X$ and~$Y$ are $(\rho,\tau)$-structured,
where
\begin{align*}
\tau & \eqdef2\cdot\delta-\varepsilon\\
\text{(by definition of\,\ensuremath{\delta})} & =2\cdot\left(1-\frac{\eta}{4}+\frac{\varepsilon}{2}\right)-\varepsilon\\
 & =2-\frac{\eta}{2}\\
 & =2+\frac{\eta}{2}-\eta\\
\text{(Since \ensuremath{\varepsilon\eqdef\frac{h}{c\cdot\eta}})} & =2+\frac{h}{2\cdot c\cdot\varepsilon}-\eta\\
 & \ge2+\frac{h'}{c\cdot\varepsilon}-\eta+\frac{1}{b},
\end{align*}
where the last inequality can be made to hold by choosing $h$ to
be sufficiently large.

Next, in Step~\enuref{deterministic-simulation-choosing-message},
the tree conditions~$X$ on the event of sending the message~$M$,
which has probability at least~$2^{-\left|M\right|}$. By \facref{min-entropy-conditioning},
this decreases $\Hm(X_{\free(\rho)})$ by at most $\left|M\right|$
bits, which increases the deficiency by at most $\left|M\right|$
bits. All in all, we showed that the first two steps of the simulation
increase the deficiency by at most~$\left|M\right|+1$.

Let $I$ be the set of queries chosen in Step~\enuref{deterministic-simulation-taking-heavy-value}.
We turn to show that the rest of the steps decrease the deficiency
by at least~$\Omega(b\cdot\left|I\right|)$. Without loss of generality,
assume that $I\ne\emptyset$ (otherwise the latter bound holds vacuously).
The rest of the steps apply the following changes to the deficiency:
\begin{itemize}
\item Step~\enuref{deterministic-simulation-taking-heavy-value} conditions~$X$
on the event $X_{I}=x_{I}$, which has probability greater than~$2^{-\delta\cdot b\cdot\left|I\right|}$
by the definition of~$x_{I}$. Hence, this conditioning increases
the deficiency by less than $\delta\cdot b\cdot\left|I\right|$ (by
\facref{min-entrpoy-projecting}).
\item Step~\enuref{deterministic-simulation-query} removes the set~$I$
from $\free(\rho)$. Looking at the definition of deficiency, this
change decreases the term of $2\cdot b\cdot\left|\free(\rho)\right|$
by~$2\cdot b\cdot\left|I\right|$, decreases the term $H_{\infty}(Y_{\free(\rho)})$
by at most $b\cdot\left|I\right|$ (by \facref{min-entrpoy-projecting}),
and does not change the term $H_{\infty}(X_{\free(\rho)})$ (since
at this point $X_{I}$ is fixed to~$x_{I}$). All in all, the deficiency
is decreased by at least~$b\cdot\left|I\right|$.
\item Finally, Step~\enuref{deterministic-simulation-consistent-with-queries}
conditions~$Y$ on the event $g^{I}(x_{I},Y_{I})=\rho_{I}$. This
event has probability at least $2^{-\left|I\right|-1}$ by the assumption
that $X$ is not dangerous (and hence not leaking). Thus, this conditioning
increases the deficiency by at most $\left|I\right|+1$ (by \facref{min-entrpoy-projecting}).
\end{itemize}
Summing all those effects together, we get that the deficiency was
decreased by at least
\[
b\cdot\left|I\right|-\delta\cdot b\cdot\left|I\right|-\left(\left|I\right|+1\right)\ge(1-\delta-\frac{2}{b})\cdot b\cdot\left|I\right|.
\]
By choosing $c$ to be sufficiently large, we can make sure that $1-\delta-\frac{2}{b}$
is a positive constant independent of $b$ and~$n$, and therefore
the decrease in the deficiency will be at least $\Omega(b\cdot\left|I\right|)$,
as required. To see it, observe that
\begin{align*}
\delta+\frac{2}{b} & =1-\frac{\eta}{4}+\frac{\varepsilon}{2}+\frac{2}{b}\\
\text{(Since \ensuremath{\varepsilon\eqdef\frac{h}{c\cdot\eta}})} & =1-\frac{\eta}{4}+\frac{h}{2\cdot c\cdot\eta}+\frac{2}{b}\\
\text{(Since \ensuremath{b\ge c})} & \le1-\frac{\eta}{4}+\frac{h+4}{2\cdot c\cdot\eta}.
\end{align*}
Therefore, if we choose $c>\frac{2\cdot(h+4)}{\eta^{2}}$, the expression
on the right-hand side will be a constant that is strictly smaller
than~$1$, as required.

\section{\label{sec:randomized-lifting}The randomized lifting theorem}

In this section, we prove the randomized part of our main theorem.
In fact, we prove the following more general result.
\begin{theorem}[Randomized lifting theorem]
\label{thm:randomized-lifting}For every $\eta>0$ there exists $c=O(\frac{1}{\eta^{2}}\cdot\log\frac{1}{\eta})$
such that the following holds: Let $n\in\N$ be such that $n\ge2$,
let $\D\eqdef\B^{b}$ be such that $b\ge c\cdot\log n$, let $g:\D\times\D\to\B$
be a function such that $\disc(g)\le2^{-\eta\cdot b}$, and let $G=g^{n}$.
Let $\Pi$ be a randomized (public-coin) protocol that takes inputs
in $\D^{n}\times\D^{n}$ that has communication complexity~$C\le2\cdot b\cdot n$
and round complexity~$r$. Then, there exists a randomized parallel
decision tree~$T$ with the following properties:
\begin{itemize}
\item On input $z\in\B^{n}$, the tree outputs a transcript~$\pi$ of~$\Pi$,
whose distribution is $2^{-\frac{\eta}{8}\cdot b}$-close to the distribution
of the transcripts of~$\Pi$ when given inputs that are uniformly
distributed in~$G^{-1}(z)$.
\item The tree $T$ has query complexity $O(\frac{C}{b}+1)$ and depth~$r$.
\end{itemize}
\end{theorem}
\noindent We first observe that \thmref{randomized-lifting} indeed
implies the lower bound of our main theorem.
\begin{myproof}[Proof of \thmref{main} from \thmref{randomized-lifting}.]
Let $\cS:\B^{n}\to\cO$ be a search problem, and let $\varepsilon>0$
and $\varepsilon'\eqdef\varepsilon+2^{-\frac{\eta}{8}\cdot b}$. We
prove that $\rcc_{\varepsilon}(\cS\circ G)=\Theta\left(\rdt_{\varepsilon'}(\cS)\cdot b\right)$.
Let $\Pi$ be an optimal protocol that solves~$\cS\circ G$ with
complexity~$C\eqdef\rcc(\cS\circ G)$, and observe that we can assume
without loss of generality that $C\le2\cdot b\cdot n$ (since the
players can solve any search problem by sending their whole inputs).
By applying the theorem to $\Pi$, we construct a tree~$T$ that
on input~$z$ samples a transcript of~$\Pi$ as in the theorem,
and outputs the output that is associated with this transcript. It
is not hard to see that the output of~$T$ will be in~$\cS(z)$
with probability at least 
\[
1-\varepsilon-2^{-\frac{\eta}{8}\cdot b}\ge1-\varepsilon',
\]
 and that the query complexity of~$T$ is $O(\frac{C}{b}+1)$. This
implies that $\rdt_{\varepsilon'}(\cS)=O\left(\rcc_{\varepsilon}(\cS\circ G)/b+1\right)$,
or in other words, $\rcc_{\varepsilon}(\cS\circ G)=\Omega\left(\left(\rdt_{\varepsilon'}(\cS)-O(1)\right)\cdot b\right)$,
as required.
\end{myproof}
In the rest of this section we prove \thmref{randomized-lifting}.
We start the proof by observing that it suffices to prove the theorem
for the special case in which the protocol~$\Pi$ is deterministic.
To see why, recall that a randomized public-coin protocol is a distribution
over deterministic protocols. Thus, if we prove the theorem for deterministic
protocols, we can extend it to randomized protocols as follows: Given
a randomized protocol~$\Pi$, the tree~$T$ will start by sampling
a deterministic protocol~$\Pi_{\mathrm{det}}$ from the distribution~$\Pi$,
and will then apply the theorem to~$\Pi_{\mathrm{det}}$. It is not
hard to verify that such a tree~$T$ satisfies the requirements of
\thmref{randomized-lifting}. Thus, it suffices to consider the case
where $\Pi$ is deterministic.

For the rest of this section, fix $\Pi$ to be an arbitrary deterministic
protocol that takes inputs in $\D^{n}\times\D^{n}$, and denote by
$C$ and~$r$ its communication complexity and round complexity respectively.
The rest of this section is organized as follows: We first describe
the construction of the parallel decision tree~$T$ in \Subsecref{randomized-construction}.
We then prove that the transcript that $T$ outputs is distributed
as required in \Subsecref{randomized-correctness}. Finally, we upper
bound the query complexity of~$T$ in \Subsecref{randomized-complexity}.

\subsection{\label{subsec:randomized-construction}The construction of~$T$}

The construction of the randomized tree~$T$ is similar to the construction
of the deterministic lifting theorem (\Subsecref{deterministic-construction}),
but has the following differences in the simulation:
\begin{itemize}
\item In the deterministic construction, the tree chose the message~$M$
arbitrarily subject to having sufficiently high probability. The reason
we could do it is that it did not matter which transcript the tree
would output as long as it was consistent in~$G^{-1}(z)$. In the
randomized construction, on the other hand, we would like to output
a transcript whose distribution is close to the ``correct'' distribution.
Therefore, we change the construction such that the message~$M$
is chosen randomly according to the distribution of the inputs.
\item Since the messages are now sampled according to the distribution of
the inputs, we can no longer guarantee that the message~$M$ has
sufficiently high probability. Therefore, the tree may choose messages~$M$
that have very low probability, and such messages may reveal too much
information about the inputs. In order to avoid that, the tree maintains
a variable~$K$ which keeps track of the amount of information that
was revealed by the messages. If at any point $K$~becomes too large,
the tree halts and declares failure. This modification is important
since if we allow the chosen messages to reveal too much information,
then they will lead the tree to make too many queries. In particular,
the bound on~$K$ is used in \subsecref{randomized-complexity} to
upper bound the query complexity of~$T$.
\item In the deterministic construction, the tree restored the density of~$X$
by fixing some set of coordinates~$I$ to some value~$x_{I}$ (using
\proref{density-restoring-fixing}). Again, this was possible since
it did not matter which transcript the tree would output. In the randomized
construction, we cannot do it, since the transcript has to be distributed
in a way that is close to be correct. In order to resolve this issue,
we follow \cite{GPW17} and use their ``density-restoring partition''
(\lemref{density-restoring-partition}). Recall that this lemma says
that the probability space of~$X$ can be partitioned into dense
parts. The tree now samples one of those parts according to their
probabilities and conditions~$X$ on being in this part. If this
conditioning reveals too much information, then the tree halts and
declares failure.
\end{itemize}
We turn to give a formal description of the construction. Let $h'$
be the maximum among the universal constants of the uniform marginals
lemma (\lemref{uniform-marginals}) and the main technical lemma (\lemref{main}),
and let $h$ be a universal constant that will be chosen to be sufficiently
large to make the inequalities in the proof hold. Let $\varepsilon\eqdef\frac{h\cdot\log c}{c\cdot\eta}$,
and as before, $\delta\eqdef1-\frac{\eta}{4}+\frac{\varepsilon}{2}$,
and $\tau\eqdef2\cdot\delta-\varepsilon$. As before, the parallel
decision tree~$T$ constructs the transcript~$\pi$ by simulating
the protocol~$\Pi$ round-by-round, each time adding a single message
to~$\pi$. Throughout the simulation, the tree maintains a rectangle~$\cX\times\cY\subseteq\D^{n}\times\D^{n}$
of inputs that are consistent with~$\pi$ (but not necessarily of
all such inputs). In what follows, we denote by $X$ and~$Y$ random
variables that are uniformly distributed over $\cX$ and~$\cY$ respectively.
As before, the tree will maintain the invariant that $X$ and~$Y$
are $(\rho,\tau)$-structured, and that moreover, they are $(\delta-\varepsilon)$-dense
and $\delta$-dense respectively in Alice's rounds and the other way
around in Bob's rounds. As mentioned above, the tree will also maintain
a variable~$K$ from iteration to iteration, which will measure the
information revealed so far.

When the tree~$T$ starts the simulation, the tree sets the transcript
$\pi$ to be the empty string, the restriction $\rho$ to~$\left\{ *\right\} ^{n}$,
the variable~$K$ to zero, and the sets $\cX,\cY$ to~$\D^{n}$.
At this point the invariant clearly holds. We now explain how $T$
simulates a single round of the protocol while maintaining the invariant.
Suppose that the invariant holds at the beginning of the current round,
and assume without loss of generality that it is Alice's turn to speak.
The tree~$T$ performs the following steps:
\begin{enumerate}
\item \label{enu:randomized-simulation-dangerous-values}The tree conditions~$X_{\free(\rho)}$
on not taking a value that is $\varepsilon$-dangerous for $Y_{\free(\rho)}$
(i.e., the tree removes from~$\cX$ all the values $x$ for which
$x_{\free(\rho)}$ is $\varepsilon$-dangerous for~$Y_{\free(\rho)}$).
\item \label{enu:randomized-simulation-choosing-message}The tree samples
a message~$M$ of Alice according to the distribution induced by~$X$.
Let $p_{M}$ be the probability of~$M$. The tree adds~$M$ to the
transcript, adds $\log\frac{1}{p_{M}}$ to~$K$, and conditions~$X$
on~$M$ (i.e., the tree sets $\cX$ to be the subset of inputs that
are consistent with~$M$).
\item \label{enu:andomized-simulation-truncating-too-much-information}If
$K>C+b$, the tree halts and declares error.
\item \label{enu:randomized-simulation-restoring -density}Let $\cX_{\free(\rho)}=\cX^{1}\cup\ldots\cup\cX^{\ell}$
be the density-restoring partition of \Lemref{density-restoring-partition}
with respect to~$X_{\free(\rho)}$. The tree chooses a random class
in the partition, where the class~$\cX^{j}$ is chosen with probability
$\Pr\left[X_{\free(\rho)}\in\cX^{j}\right]$. Let $\cX^{j}$ be the
chosen class, and let $I_{j}$ and $x_{I_{j}}$ be the set and the
value associated with~$\cX^{j}$. The tree conditions $X$ on the
event $X_{\free(\rho)}\in\cX^{j}$ (i.e., the tree sets~$\cX$ to
be the subset of inputs~$x$ such that $x_{\free(\rho)}\in\cX^{j}$).
The variable $X_{\free(\rho)-I_{j}}$ is now $\delta$-dense by the
properties of the density-restoring partition.
\item \label{enu:randomized-simulation-truncating-small-classes}Recall
that
\[
p_{\ge j}\eqdef\Pr\left[X_{\free(\rho)}\in\cX^{j}\cup\ldots\cup\cX^{\ell}\right],
\]
(see \Lemref{density-restoring-partition}). If $p_{\ge j}<\frac{1}{8}\cdot2^{-\frac{\eta}{8}}\cdot\frac{1}{2\cdot n\cdot b}$,
the tree halts and declares error.
\item \label{enu:randomized-simulation-query}The tree queries the coordinates
in~$I_{j}$, and updates~$\rho$ accordingly.
\item \label{enu:randomized-simulation-consistent-with-queries}The tree
conditions $Y$~on $g^{I}(x_{I_{j}},Y_{I_{j}})=\rho_{I_{j}}$ (i.e.,
the tree sets $\cY$ to be the subset of values~$y$ for which $g^{I}(x_{I_{j}},Y_{I_{j}})=\rho_{I_{j}}$).
Due to Step~\enuref{deterministic-simulation-dangerous-values},
the variable~$X_{\free(\rho)}$ must take a value that is not $\varepsilon$-dangerous,
and therefore $Y_{\free(\rho)}$ is necessarily $(\delta-\varepsilon)$-dense.
\end{enumerate}
After those steps take place, it becomes Bob's turn to speak, and
indeed, $X_{\free(\rho)}$ and $Y_{\free(\rho)}$ are $\delta$-dense
and $(\delta-\varepsilon)$-dense respectively. Thus, the invariant
is maintained. When the protocol~$\Pi$ stops, the tree~$T$ outputs
the transcript~$\pi$ and halts. The proof that the above steps are
well-defined is similar to the proof for the deterministic construction
and is therefore omitted.

\paragraph*{The depth of~$T$.}

As in the proof of the deterministic lifting theorem, it is not hard
to see that the depth of~$T$ is equal to the round complexity of~$\Pi$.

\subsection{\label{subsec:randomized-correctness}The correctness of~$T$}

In this section, we prove the correctness of the construction. For
convenience, we first prove the correctness of a modified tree~$T^{*}$,
whose construction is the same as that of~$T$ except that Step~\ref{enu:andomized-simulation-truncating-too-much-information}
is omitted. Fix an input $z\in\B^{n}$. We define the following (random)
transcripts of the protocol~$\Pi$:
\begin{itemize}
\item Let $\pi$ be a transcript that $T$ outputs when given~$z$.
\item Let $\pi^{*}$ be a transcript that $T^{*}$ outputs when given~$z$.
\item Let $\pi'$ be a transcript of~$\Pi$ when given inputs $(X',Y')$
that are uniformly distributed in~$G^{-1}(z)$.
\end{itemize}
Our end goal is to prove that $\pi$ and~$\pi'$ are $2^{-\frac{\eta}{8}\cdot b}$-close.
In order to do so, we will first prove that $\pi^{*}$ is $(\frac{1}{2}\cdot2^{-\frac{\eta}{8}\cdot b})$-close
to $\pi'$. We will then prove that $\pi$ is $2^{-b}$-close to $\pi^{*}$.
Together, the two results imply that $\pi$ is $2^{-\frac{\eta}{8}\cdot b}$-close
to $\pi'$, as required.

\paragraph*{$\pi^{*}$ is close to $\pi'$.}

We first prove that $\pi^{*}$ is $(\frac{1}{2}\cdot2^{-\frac{\eta}{8}\cdot b})$-close
to~$\pi'$. To this end, we construct a coupling of $\pi^{*}$ and~$\pi'$
such that $\Pr\left[\pi^{*}\ne\pi\right]\le\frac{1}{2}\cdot2^{-\frac{\eta}{8}\cdot b}$.
Essentially, we construct the coupling by going over the simulation
step-by-step and using the uniform marginals lemma to argue that at
each step, $X$ and $X'$ are close and can therefore be coupled (and
similarly for $Y$ and~$Y'$). We start by setting some notation:
for every $i\in\left[r\right]$, let us denote by $\cX_{i}\times\cY_{i}$
be the rectangle $\cX\times\cY$ from \subsecref{randomized-construction}
at the end of the $i$-th round of the simulation of~$T^{*}$ (if
$T^{*}$ halts before the $i$-th round ends, set $\cX_{i}\times\cY_{i}$
to be the rectangle $\cX\times\cY$ at the end of the simulation).
In our proof, we construct, for every $i\in\left[r\right]$:
\begin{itemize}
\item A random rectangle $\cX_{i}'\times\cY_{i}'$ that is jointly distributed
with $X',Y'$ with the following property: conditioned on a specific
choice of $\cX_{i}'\times\cY_{i}'$, the pair $(X',Y')$ is uniformly
distributed over $(\cX_{i}'\times\cY_{i}')\cap G^{-1}(z)$.
\item A coupling of $\cX_{i}\times\cY_{i}$ and $\cX_{i}'\times\cY_{i}'$
such that $\Pr\left[\cX_{i}\times\cY_{i}\ne\cX_{i}'\times\cY_{i}'\right]\le\frac{1}{2}\cdot2^{-\frac{\eta}{8}\cdot b}\cdot\frac{i}{2\cdot n\cdot b}$.
\end{itemize}
Observe that if we can construct such rectangles and couplings, then
it follows that $\pi^{*}$ and~$\pi'$ are close. To see it, observe
that at any given point during the simulation, all the inputs in the
rectangle $\cX\times\cY$ are consistent with the transcript~$\pi$.
Hence, if $\cX_{r}\times\cY_{r}=\cX_{r}'\times\cY_{r}'$, it necessarily
means that the inputs $(X',Y')$ are consistent with the transcript~$\pi$,
so $\pi=\pi'$. It follows that
\begin{align*}
\Pr\left[\pi\ne\pi'\right] & \le\Pr\left[\cX_{r}\times\cY_{r}\ne\cX_{r}'\times\cY_{r}'\right]\\
 & \le\frac{1}{2}\cdot2^{-\frac{\eta}{8}\cdot b}\cdot\frac{r}{2\cdot n\cdot b}\\
 & \le\frac{1}{2}\cdot2^{-\frac{\eta}{8}\cdot b}\cdot\frac{C}{2\cdot n\cdot b}\\
 & \le\frac{1}{2}\cdot2^{-\frac{\eta}{8}\cdot b},
\end{align*}
as required.

It remains to construct the rectangles $\cX_{i}'\times\cY_{i}'$ and
the associated couplings. We construct them by induction. Let $i\in\left[r\right]$,
and suppose we have already constructed $\cX_{i-1}'\times\cY_{i-1}'$
and its associated coupling (here, if $i=1$ we set both $\cX_{i-1}\times\cY_{i-1}$
and $\cX_{i-1}'\times\cY_{i-1}'$ to $\D^{n}\times\D^{n}$). The $i$-th
coupling first samples $\cX_{i-1}\times\cY_{i-1}$ and $\cX_{i-1}'\times\cY_{i-1}'$
from the $(i-1)$-th coupling. If they are different, then we set
$\cX_{i}'\times\cY_{i}'$ arbitrarily and assume that the coupling
failed (i.e., $\cX_{i}\times\cY_{i}$ and $\cX_{i}'\times\cY_{i}'$
are different). Suppose now that $\cX_{i-1}\times\cY_{i-1}$ and $\cX_{i-1}'\times\cY_{i-1}'$
are equal, and condition on some specific choice of this rectangle.
If the tree~$T^{*}$ has already halted by this point, we set $\cX_{i}'\times\cY_{i}'=\cX_{i-1}'\times\cY_{i-1}'$.
Otherwise, we proceed as follows.

Let $(X,Y)$ be a random pair that is uniformly distributed over $\cX_{i-1}\times\cY_{i-1}$,
and recall that due to our conditioning, the pair $(X',Y')$ is uniformly
distributed over $(\cX_{i-1}'\times\cY_{i-1}')\cap G^{-1}(z)$. We
construct the rest of the coupling by following the simulation step-by-step.
For Step~\ref{enu:randomized-simulation-dangerous-values}, with
probability
\[
\Pr\left[X_{\free(\rho)}'\text{ is \ensuremath{\varepsilon}-dangerous for }Y_{\free(\rho)}\right]
\]
we assume that the coupling failed and set $\cX_{i}'\times\cY_{i}'$
arbitrarily. Otherwise, we condition both $X$ and $X'$ on not taking
a dangerous value. In order to analyze the probability of failure,
recall that at the beginning of this step, $(X,Y)$ are $(\rho,\tau)$-structured,
where
\begin{align*}
\tau & \eqdef2\cdot\delta-\varepsilon\\
\text{(by definition of\,\ensuremath{\delta})} & =2\cdot\left(1-\frac{\eta}{4}+\frac{\varepsilon}{2}\right)-\varepsilon\\
 & =2-\frac{\eta}{2}\\
 & \ge2+\frac{\eta}{4}-\eta+\frac{\eta}{8}\\
\text{(Since \ensuremath{\varepsilon\eqdef\frac{h\cdot\log c}{c\cdot\eta}})} & =2+\frac{h\cdot\log c}{4\cdot c\cdot\varepsilon}-\eta+\frac{\eta}{8}\\
 & \ge2+\frac{h'}{c\cdot\varepsilon}-\eta+\frac{\eta}{8}+\frac{3\log c}{c}+\frac{4}{b},
\end{align*}
where the last inequality can be made to hold by choosing $h$ to
be sufficiently large. Hence, our main technical lemma (\lemref{main})
implies that the probability that $X_{\free(\rho)}$ is $\varepsilon$-dangerous
for $Y_{\free(\rho)}$ is at most 
\[
2^{-(\frac{\eta}{8}+\frac{3\log c}{c}+\frac{4}{b})\cdot b}\le\frac{1}{8}\cdot2^{-\frac{\eta}{8}\cdot b}\cdot\frac{1}{2\cdot n\cdot b}.
\]
Moreover, the uniform marginals lemma (\lemref{uniform-marginals})
implies that $X'$ is $\left(\frac{1}{8}\cdot2^{-\frac{\eta}{8}\cdot b}\cdot\frac{1}{2\cdot n\cdot b}\right)$-close
to $X$ and therefore the probability that $X_{\free(\rho)}'$ is
$\varepsilon$-dangerous for $Y_{\free(\rho)}$ is at most~$2\cdot\frac{1}{8}\cdot2^{-\frac{\eta}{8}\cdot b}\cdot\frac{1}{2\cdot n\cdot b}$.
Hence, the failure probability at this step is at most~$\frac{1}{4}\cdot2^{-\frac{\eta}{8}\cdot b}\cdot\frac{1}{2\cdot n\cdot b}$.
Note that if the coupling does not fail, $X$ is conditioned on an
event of probability at least $\frac{1}{2}$, and therefore after
the conditioning $X$ and~$Y$ are $(\rho,\tau-\frac{1}{b})$-structured.

For Steps \ref{enu:randomized-simulation-choosing-message} and~\enuref{randomized-simulation-restoring -density},
let $M$ and $\cX^{j}$ be the message and partition class that are
distributed according to the input $X$ respectively. Let $M'$ and
${\cX^{j}}'$ be the corresponding message and class of~$X'$, Since
$X$ and~$Y$ are $(\rho,\tau-\frac{1}{b})$-structured, it can again
be showed by the uniform marginals lemma that $X$ and $X'$ are $\left(\frac{1}{8}\cdot2^{-\frac{\eta}{8}\cdot b}\cdot\frac{1}{2\cdot n\cdot b}\right)$-close,
and therefore the pair $(M,\cX^{j})$ is $\left(\frac{1}{8}\cdot2^{-\frac{\eta}{8}\cdot b}\cdot\frac{1}{2\cdot n\cdot b}\right)$-close
to the pair $(M',{\cX^{j}}')$. This implies that there exists a coupling
of $(M,\cX^{j})$ and~$(M',{\cX^{j}}')$ such that the probability
that they differ is at most $\frac{1}{8}\cdot2^{-\frac{\eta}{8}\cdot b}\cdot\frac{1}{2\cdot n\cdot b}$.
We sample $(M,\cX^{j})$ and~$(M',{\cX^{j}}')$ from this coupling.
If they differ, we assume that the coupling failed, and set $\cX_{i}'\times\cY_{i}'$
arbitrarily. Otherwise, we condition both $X$ and $X'$ on being
consistent with the message~$M$ and the class $\cX^{j}$, and denote
by $I_{j},x_{I_{j}}$ the set and values associated with~$\cX^{j}$.
Finally, for Step~\ref{enu:randomized-simulation-truncating-small-classes},
if $p_{\ge j}\le\frac{1}{8}\cdot2^{-\frac{\eta}{8}\cdot b}\cdot\frac{1}{2\cdot n\cdot b}$,
then we assume that the coupling fails and set $\cX_{i}'\times\cY_{i}'$
arbitrarily (note that this happens with probability at most $\frac{1}{8}\cdot2^{-\frac{\eta}{8}\cdot b}\cdot\frac{1}{n\cdot b}$).

At this point, we set $\cX_{i}'=\cX^{j}$, and set $\cY_{i}'$ to
be the set of inputs $y\in\cY_{i-1}$ for which $g(x_{I_{j}},y_{I_{j}})=z_{I_{j}}$.
It is easy to see that this choice satisfies $\cX_{i}'\times\cY_{i}'=\cX_{i}\times\cY_{i}$.
To analyze the total failure probability of this coupling, observe
that by the induction assumption, the failure probability of the $(i-1)$-th
coupling is at most $\frac{1}{2}\cdot2^{-\frac{\eta}{8}\cdot b}\cdot\frac{i-1}{2\cdot n\cdot b}$,
and the other failure events discussed above at to that a failure
probability of at most 
\[
\left(\frac{1}{4}+\frac{1}{8}+\frac{1}{8}\right)\cdot2^{-\frac{\eta}{8}\cdot b}\cdot\frac{1}{2\cdot n\cdot b}=\frac{1}{2}\cdot2^{-\frac{\eta}{8}\cdot b}\cdot\frac{1}{2\cdot n\cdot b}.
\]
Hence, the failure probability of the $i$-th coupling is at most
$\frac{3}{4}\cdot2^{-\frac{\eta}{4}\cdot b}\cdot\frac{i}{n\cdot b}$,
as required.

It remains to show that conditioned on any specific choice of $\cX_{i}'\times\cY_{i}'$,
the pair $(X',Y')$ is uniformly distributed over $(\cX_{i}'\times\cY_{i}')\cap G^{-1}(z)$.
In the cases where the coupling fails, we can ensure this property
holds by first sampling $(X',Y')$ and then setting $\cX_{i}'\times\cY_{i}'=\left\{ (X',Y')\right\} $.
Suppose that the coupling did not fail. Recall that by the induction
assumption, it holds that conditioned on the choice of $\cX_{i-1}'\times\cY_{i-1}'$,
the pair $(X',Y')$ is uniformly distributed over $(\cX_{i-1}'\times\cY_{i-1}')\cap G^{-1}(z)$.
Observe that all the $i$-th coupling changes in the distribution
of $(X',Y')$ is to condition it on being in $\cX_{i}'\times\cY_{i}'$.
Thus, at the end of the $i$-th coupling, the pair $(X',Y')$ is uniformly
distributed over $(\cX_{i}'\times\cY_{i}')\cap G^{-1}(z)$, as required.

\paragraph*{$\pi$ is close to $\pi^{*}$.}

We turn to prove that $\pi$ is $2^{-b}$-close to $\pi^{*}$. Let
$\cE$ denote the event that the tree~$T$ halts in Step~\ref{enu:andomized-simulation-truncating-too-much-information}.
It is not hard to see that the statistical distance between $\pi$
and~$\pi^{*}$ is exactly $\Pr\left[\cE\right]$. We show that $\Pr\left[\cE\right]<2^{-b}$,
and this will conclude the proof of correctness.

Intuitively, the reason that $\Pr\left[\cE\right]<2^{-b}$ is that
the tree halts only if the probability of the transcript up to that
point is less than $2^{-C-b}$: to see it, observe that the variable
$K$ measures (roughly) the logarithm of the probability of the transcript
up to that point, and recall that the tree halts when $K>C+b$. By
taking union bound over all possible transcripts, we get that the
halting probability is less than~$2^{-b}$.

Unfortunately, the formal proof contains a messier calculation: the
reason is that the probabilities of the messages as measured by~$K$
depend on the choices of the classes~$\cX^{j}$ in Step~\ref{enu:randomized-simulation-restoring -density},
so the foregoing intuition only holds for a given choice of these
classes. Thus, the formal proof also sums over all the possible choices
of classes~$\cX^{j}$ and conditions on those choices. However, while
the resulting calculation is more complicated, the idea is the same.

In order to facilitate the formal proof, we setup some useful notation.
Let $M_{1},\ldots,M_{r}$ be the messages that are chosen in Step~\ref{enu:randomized-simulation-choosing-message}
of the simulation (so $\pi=(M_{1},\ldots,M_{r})$), and let $J=(j_{1},\ldots,j_{r})$
be the indices of the classes that are chosen in Step~\ref{enu:randomized-simulation-restoring -density}
(if the tree halts before the $i$-th round, set $M_{i}$ to the empty
string and set $j_{i}=1$). Observe that the execution of~$T$ is
completely determined by $\pi$ and~$J$, and in particular, $\pi$
and~$J$ determine whether the event~$\cE$ happens or not. With
some abuse of notation, let us denote the fact that a particular choice
of $(\pi,J)$ is consistent with~$\cE$ by $(\pi,J)\in\cE$. For
any $i\in\left[r\right]$, let us denote $\pi_{\le i}=(M_{1},\ldots,M_{i-1})$
and $J_{\le i}=(j_{1},\ldots,j_{i})$. Observe that at the $i$-th
round, the probability $p_{M_{i}}$ in Step~\ref{enu:randomized-simulation-choosing-message}
is determined by $\pi_{<i}$ and $J_{<i}$, and let us denote by $p_{M_{i}\texttt{\ensuremath{\mid}}\pi_{<i},J_{<i}}$
this probability for a given choice of $\pi_{<i}$ and $J_{<i}$.
We are now ready to prove the upper bound on~$\Pr\left[\cE\right]$.
It holds that
\begin{align*}
\Pr\left[\cE\right] & =\sum_{(\pi,J)\in\cE}\Pr\left[\pi,J\right]\\
 & =\sum_{(\pi,J)\in\cE}\Pr\left[M_{1}\right]\cdot\Pr\left[j_{1}\texttt{\ensuremath{\mid}}\pi_{\le1}\right]\cdots\Pr\left[M_{r}\texttt{\ensuremath{\mid}}\pi_{\le r-1},J_{\le r-1}\right]\cdot\Pr\left[j_{r}\texttt{\ensuremath{\mid}}\pi_{\le r},J_{\le r-1}\right]\\
 & =\sum_{(\pi,J)\in\cE}p_{M_{1}}\cdot p_{M_{2}\texttt{\ensuremath{\mid}}\pi_{\le1},J_{\le1}}\cdots p_{M_{r}\texttt{\ensuremath{\mid}}\pi_{\le r-1},J_{\le r-1}}\\
 & \,\,\,\,\,\,\,\cdot\Pr\left[j_{1}\texttt{\ensuremath{\mid}}\pi_{\le1}\right]\cdots\Pr\left[j_{r}\texttt{\ensuremath{\mid}}\pi_{\le r},J_{\le r-1}\right].
\end{align*}
Next, observe that for every choice of $(\pi,J)$, the corresponding
value of~$K$ at the end of the simulation is
\[
\log\frac{1}{p_{M_{1}}}+\log\frac{1}{p_{M_{2}\texttt{\ensuremath{\mid}}\pi_{<2},J_{<2}}}+\ldots+\log\frac{1}{p_{M_{r}\texttt{\ensuremath{\mid}}\pi_{<r},J_{<r}}}.
\]
In particular, if $(\pi,J)\in\cE$, then it holds that $K>C+b$, and
therefore
\[
p_{M_{1}}\cdot p_{M_{2}\texttt{\ensuremath{\mid}}\pi_{<2},J_{<2}}\cdots p_{M_{r}\texttt{\ensuremath{\mid}}\pi_{<r},J_{<r}}<2^{-C-b}.
\]
It follows that
\begin{align}
\Pr\left[\cE\right] & =\sum_{(\pi,J)\in\cE}p_{M_{1}}\cdot p_{M_{2}\texttt{\ensuremath{\mid}}\pi_{<2},J_{<2}}\cdots p_{M_{r}\texttt{\ensuremath{\mid}}\pi_{<r},J_{<r}}\nonumber \\
 & \,\,\,\,\,\,\,\cdot\Pr\left[j_{1}\texttt{\ensuremath{\mid}}\pi_{\le1}\right]\cdots\Pr\left[j_{r}\texttt{\ensuremath{\mid}}\pi_{\le r},J_{\le r-1}\right]\nonumber \\
 & <\sum_{(\pi,J)\in\cE}2^{-C-b}\cdot\Pr\left[j_{1}\texttt{\ensuremath{\mid}}\pi_{\le1}\right]\cdots\Pr\left[j_{r}\texttt{\ensuremath{\mid}}\pi_{\le r},J_{\le r-1}\right]\nonumber \\
 & \le\sum_{(\pi,J)}2^{-C-b}\cdot\Pr\left[j_{1}\texttt{\ensuremath{\mid}}\pi_{\le1}\right]\cdots\Pr\left[j_{r}\texttt{\ensuremath{\mid}}\pi_{\le r},J_{\le r-1}\right]\nonumber \\
 & \le2^{-C-b}\cdot\sum_{M_{1},j_{1}}\Pr\left[j_{1}\texttt{\ensuremath{\mid}}\pi_{\le1}\right]\cdot\sum_{M_{2},j_{2}}\Pr\left[j_{2}\texttt{\ensuremath{\mid}}\pi_{\le2},J_{\le1}\right]\nonumber \\
 & \,\,\,\,\,\,\,\,\,\,\,\,\,\,\,\,\,\,\,\cdots\sum_{M_{r},j_{r}}\Pr\left[j_{2}\texttt{\ensuremath{\mid}}\pi_{\le2},J_{\le1}\right]\nonumber \\
 & =2^{-C-b}\cdot\sum_{M_{1}}1\cdot\sum_{M_{2}}1\cdots\sum_{M_{r}}1\label{eq:total-probability}\\
 & =2^{-C-b}\cdot\sum_{\pi}1\nonumber \\
 & \le2^{-b},\label{eq:number-of-transcripts}
\end{align}
as required. In the calculation above, Equality~\ref{eq:total-probability}
follows since each sum goes over all the possible choices of~$j_{i}$,
and Inequality~\ref{eq:number-of-transcripts} follows since $\Pi$
has at most $2^{C}$~distinct transcripts.

\subsection{\label{subsec:randomized-complexity}The query complexity of~$T$}

The analysis of the query complexity here is similar to the analysis
of the deterministic query complexity. The main difference is the
following: In the deterministic setting, the increase in the deficiency
due to a single message~$M$ was upper bounded by~$\left|M\right|$,
and therefore the total increase in the deficiency was upper bounded
by~$\left|C\right|$. In the randomized case, the increase in the
deficiency due to a single message~$M$ is upper bounded by $\log\frac{1}{p_{M}}$.
Thus, we upper bound the total increase in the deficiency by~$K$.
Since $K$ is never larger than $C+b$ due to Step~\enuref{andomized-simulation-truncating-too-much-information},
we conclude that the query complexity is at most $O(\frac{C+b}{b})=O(\frac{C}{b}+1)$.
Details follow.

As before, we define the deficiency of~$X,Y$ to be 
\[
2\cdot b\cdot\left|\free(\rho)\right|-H_{\infty}(X_{\free(\rho)})-H_{\infty}(Y_{\free(\rho)}).
\]
We prove that whenever the protocol transmits a message~$M$, the
deficiency increases by~$O(\log\frac{1}{p_{M}})$, and that whenever
the tree~$T$ makes a query, the deficiency is decreased by~$\Omega(b)$.
Since the deficiency is always non-negative, and $K$ is never more
than $C+b$, it will follow that the tree must make at most $O(\frac{C+b}{b})$
queries. More specifically, we prove that in every round, the first
two steps increase the deficiency by $\log\frac{1}{p_{M}}+1$, and
the rest of the steps decrease the deficiency by $\Omega(\left|I_{j}\right|\cdot b)$,
and this will imply the desired result.

Fix a round of the simulation, and assume without loss of generality
that the message is sent by Alice. We start by analyzing Step~\enuref{randomized-simulation-dangerous-values}.
At this step, the tree conditions~$X_{\free(\rho)}$ on taking dangerous
values that are not $\varepsilon$-dangerous for~$Y_{\free(\rho)}$.
Using the same calculation as in \Subsecref{randomized-correctness},
it can be showed that the probability of non-dangerous values is at
least~$\frac{1}{2}$. Therefore, this step increases the deficiency
by at most $1$~bit. Next, in Step~\enuref{randomized-simulation-choosing-message},
the tree conditions~$X$ on an event of choosing the message~$M$,
whose probability is $p_{M}$ by definition. Thus, this step increases
the deficiency by at most $\log\frac{1}{p_{M}}$ bits. All in all,
we showed that the first two steps of the simulation increase the
deficiency by at most~$\log\frac{1}{p_{M}}+1$ bits.

Let $\cX^{j}$ be the partition class that is sampled in Step~\enuref{randomized-simulation-restoring -density},
and let $I_{j},x_{j}$ be the set and value that are associated with
$\cX^{j}$. We turn to show that the rest of the steps decrease the
deficiency by $\Omega(b\cdot\left|I_{j}\right|)$. Those steps apply
the following changes to the deficiency:
\begin{itemize}
\item Step~\enuref{randomized-simulation-restoring -density} conditions~$X$
on the event $X_{\free(\rho)}\in\cX^{j}$. By \lemref{density-restoring-partition},
this conditioning increases the deficiency at most $\delta\cdot b\cdot\left|I\right|+\log\frac{1}{p_{\ge j}}$.
Recall that by Step~\enuref{randomized-simulation-truncating-small-classes},
the probability $p_{\ge j}$ can never be less than $\frac{1}{8}\cdot2^{-\frac{\eta}{8}\cdot b}\cdot\frac{1}{2\cdot n\cdot b}$.
Thus, this step increases the deficiency by at most
\[
\delta\cdot b\cdot\left|I\right|+\frac{\eta}{8}\cdot b+\log(2\cdot n\cdot b)+3\le(\delta+\frac{\eta}{4}+\frac{6}{c})\cdot b\cdot\left|I\right|.
\]
\item Step \enuref{randomized-simulation-query} removes the set~$I$ from
$\free(\rho)$. Looking at the definition of deficiency, this change
decreases the term of $2\cdot b\cdot\left|\free(\rho)\right|$ by~$2\cdot b\cdot\left|I\right|$,
decreases the term $H_{\infty}(Y_{\free(\rho)})$ by at most $b\cdot\left|I\right|$
(\facref{min-entrpoy-projecting}), and does not change the term $H_{\infty}(X_{\free(\rho)})$
(since at this point $X_{I}$ is fixed to~$x_{I}$). All in all,
the deficiency is decreased by at least~$b\cdot\left|I\right|$.
\item Finally, Step~\enuref{randomized-simulation-consistent-with-queries}
conditions~$Y$ on the event $g^{I}(x_{I},Y_{I})=\rho_{I}$. This
event has probability at least $2^{-\left|I\right|-1}$ by the assumption
that $X$ is not dangerous (and hence not leaking). Thus, this conditioning
increases the deficiency by at most $\left|I\right|+1$.
\end{itemize}
Summing all those effects together, we get that the deficiency was
decreased by at least
\[
b\cdot\left|I\right|-(\delta+\frac{\eta}{8}+\frac{6}{c})\cdot b\cdot\left|I\right|-\left(\left|I\right|+1\right)\ge(1-\delta-\frac{\eta}{8}-\frac{7}{c})\cdot b\cdot\left|I\right|.
\]
By choosing $c$ to be sufficiently large, we can make sure that $1-\delta-\frac{\eta}{8}-\frac{7}{c}$
is a positive constant independent of $b$ and~$n$, and therefore
the decrease in the deficiency will be at least $\Omega(b\cdot\left|I\right|)$,
as required. To see it, observe that
\begin{align*}
\delta+\frac{\eta}{8}+\frac{7}{c} & =1-\frac{\eta}{8}+\frac{\varepsilon}{2}+\frac{\eta}{8}+\frac{7}{c}\\
\text{(Since \ensuremath{\varepsilon\eqdef\frac{h\cdot\log c}{c\cdot\eta}})} & =1-\frac{\eta}{8}+\frac{h\cdot\log c}{2\cdot c\cdot\eta}+\frac{7}{c}\\
 & \le1-\frac{\eta}{8}+\frac{(h+14)\cdot\log c}{2\cdot c\cdot\eta}.
\end{align*}
Thus, if we choose $c$ such that $\frac{c}{\log c}>\frac{h+14}{2\cdot\eta^{2}}$,
the expression on the right-hand side will be a constant that is strictly
smaller than~$1$. It is not hard to see that we can choose such
a value of~$c$ that satisfies $c=O(\frac{1}{\eta^{2}}\cdot\log\frac{1}{\eta})$.
\begin{acknowledgement*}
We thank Daniel Kane for some very enlightening conversations and
suggestions. The authors would also like to thank anonymous referees
for comments that improved the presentation of this work. Part of
this work was carried out while the authors were visiting the Simons
Institute for the Theory of Computing.
\end{acknowledgement*}

\newcommand{\etalchar}[1]{$^{#1}$}

\end{document}